\documentclass[aps,prev,twocolumn,preprintnumbers,floatfix,nofootinbib]{revtex4-1}
\pdfoutput=1
\usepackage{graphicx}
\usepackage{bm}
\usepackage{times}
\usepackage{slashed}
\usepackage{color}
\usepackage{slashed}
\usepackage{amsmath}
\usepackage{amsthm}
\usepackage{subfigure}
\usepackage{hyperref}

\newcommand{\be}{\begin{equation}}
\newcommand{\ee}{\end{equation}}
\newcommand{\bea}{\begin{eqnarray}}
\newcommand{\eea}{\end{eqnarray}}

\begin{document}

\title{Vector-Like Leptons and Inert Scalar Triplet: Lepton Flavor Violation, $g-2$ and Collider Searches}

\author{A. S. de Jesus$^{a}$}
\email{alvarosdj@ufrn.edu.br}

\author{S. Kovalenko$^{b}$}
\email{sergey.kovalenko@unab.cl}

\author{F. S. Queiroz$^{a}$}
\email{farinaldo.queiroz@iip.ufrn.br}

\author{C. Siqueira$^{a}$}
\email{csiqueira@iip.ufrn.br}

\author{K. Sinha$^{c}$}
\email{kuver.sinha@ou.edu}

\affiliation{$^a$International Institute of Physics, Universidade Federal do Rio Grande do Norte,
Campus Universit\'ario, Lagoa Nova, Natal-RN 59078-970, Brazil\\
$^b$ Departamento de Ciencias F\'isicas,
Universidad Andres Bello, Sazi\'e 2212, Santiago, Chile\\
$^c$ Department of Physics and Astronomy, University of Oklahoma, Norman, OK 73019, USA
}

\begin{abstract}

 We investigate simplified models involving an inert scalar triplet and vector-like  leptons  that can account for the muon $g-2$ anomaly. These simplified scenarios are embedded in a model that features W' and Z' bosons, which are subject to stringent collider bounds. The constraints coming from the muon $g-2$ anomaly are put into perspective with collider bounds, as well as bounds coming from lepton flavor violation searches. The region of parameter space that explains the $g-2$ anomaly is shown to be within reach of lepton flavor violation probes and future colliders such as HL-LHC and HE-LHC.
\end{abstract}


\maketitle

\section{Introduction}
\label{intro}

The Dirac Equation predicts that the muon has a  magnetic moment equal to $\vec{m}=  g e/(2 m_\mu) \vec{S}$, where $g=2$ is the gyromagnetic ratio. Quantum corrections to the $g$-factor are parametrized by the muon anomalous magnetic moment ($g-2$) defined as, 
\begin{equation}
    a_{\mu}=\frac{g-2}{2}.
\end{equation}   
Theoretical calculations of the Standard Model (SM) contributions to $g-2$ represent a remarkable success of quantum field theory. Since $a_{\mu}$ is the target of increasingly precise theoretical prediction on the one hand and vigorous experimental measurement on the other, it serves as a golden channel by which the SM can be tested at quantum loop level. Any deviation between theory and experiment would imply the existence of new physics.

Indeed, such a discrepancy between theory and experiment has been found and has persisted over many years \cite{Bennett:2002jb,Bennett:2006fi}. The precise extent of the $g-2$ anomaly may be mitigated or exacerbated by the still significant theoretical uncertainties \cite{Crivellin:2020zul,Borsanyi:2020mff}. For example, the $g-2$ anomaly goes from $3.3\sigma$ up to $5\sigma$ depending on the hadronic corrections: 
\begin{eqnarray}
    \Delta a_\mu & = & (261 \pm 78)\times 10^{-11}\,\, (3.3\sigma)\,\, \text{ \cite{Prades:2009tw,Tanabashi:2018oca} - (2009)}; \nonumber\\
     \Delta a_\mu & = &  (325 \pm 80)\times 10^{-11}\,\, (4.05\sigma)\,\, \text{ \cite{Benayoun:2012wc} - (2012)};\nonumber\\
     \Delta a_\mu & = & (287 \pm 80)\times 10^{-11}\,\, (3.6\sigma)\,\, \text{ \cite{Blum:2013xva} - (2013)};\nonumber\\
     \Delta a_\mu & = &  (377 \pm 75)\times 10^{-11}\,\, (5.02\sigma)\,\, \text{ \cite{Benayoun:2015gxa} - (2015)};\nonumber\\
          \Delta a_\mu & = &  (313 \pm 77)\times 10^{-11}\,\, (4.1\sigma)\,\, \text{ \cite{Jegerlehner:2017lbd}- (2017)};\nonumber\\
     \Delta a_\mu & = &  (270 \pm 36)\times 10^{-11}\,\, (3.7\sigma)\,\, \text{ \cite{Keshavarzi:2018mgv} - (2018)}.
     \label{deltasigmavalues}
\end{eqnarray}
The current value adopted by the Particle Data Group (PDG) is $\Delta a_\mu=(261 \pm 78)\times 10^{-11}$ \cite{Tanabashi:2018oca}, which is the value used in our work. 


Highly anticipated experimental results from FERMILAB \cite{Grange:2015fou}, \cite{Kronfeld:2013uoa} and the J-PARC \cite{Abe:2019thb} facility in Japan will further shed light on this anomaly. Recent theoretical progress in conjunction with the  anticipated increase in experimental precision holds the promise of giving a unique window into beyond SM physics in the near future. It is thus extremely important to explore new physics scenarios which may potentially explain the $g-2$ anomaly, if it does indeed become stronger.

 



Many scenarios to address the muon $g-2$ anomaly have been put forward in the context of simplified models and supersymmetry \cite{Padley:2015uma,Yamaguchi:2016oqz,Yin:2016shg,Endo:2019bcj,  Endo:2020mqz, Endo:2019mxw, Cox:2018vsv,  Konar:2017oah, Un:2016hji, An:2015uwa, Chakraborty:2015bsk,Jana:2020pxx}. We refer to \cite{Lindner:2016bgg} and references therein for a recent overview of the  status of this vast literature.

In this work, we discuss simplified models for the $g-2$ anomaly  containing  vector-like leptons and an inert scalar triplet, and explore correlations with lepton flavor violation and constraints coming from collider physics. For earlier work on vector-like leptons in the context of $g-2$, we refer to \cite{Dermisek:2013gta}, \cite{Poh:2017tfo}, \cite{Barman:2018jhz}. We embed our simplified scenarios into a model with the extended gauge group structure $SU(3)_C \times SU(3)_L \times U(1)_N$ (3-3-1 for short) \cite{Pisano:1991ee,Foot:1992rh,Hoang:1995vq} which has been widely studied in the literature. Our focus is twofold: firstly, to study the confluence of diverse experimental constraints on our simplified models and their capacity to explain the $g-2$ anomaly; and secondly, to assess the possibility of addressing the anomaly within the broader architecture of the 3-3-1 models.

We briefly describe these two aspects of our work. Firstly, the confluence of various experimental results, especially those coming from lepton flavor violation, turns out to be quite restrictive on the prospects of addressing the $g-2$ anomaly within our simplified scenarios.  We focus in particular on the rare muon decay $\mu \to e\gamma$, on which the MEG collaboration \cite{TheMEG:2016wtm} currently imposes the bound $BR(\mu \to e\gamma) < 4.2 \times 10^{-13}$. There is an ongoing effort to push this bound down to $4\times 10^{-14}$ \cite{Mori:2016vwi}. Using the bound from the MEG collaboration restricts large portions of parameter space where the $g-2$ anomaly can be addressed; this can be seen from our results in FIG. 4, FIG. 5, and FIG. 6. 

Secondly, our work is also relevant for attempts to address the $g-2$ anomaly within 3-3-1 models, since stringent collider bounds severely constrain such efforts in most avatars of these scenarios. The masses of the $W^{\prime}$ and $Z^{\prime}$ bosons in these models are proportional to the energy scale at which $SU(3)_L \times U(1)_N \rightarrow SU(2)_L \times U(1)_Y$. The LHC places strong lower bounds on the masses of these gauge bosons and typically, one finds that the energy scale of 3-3-1 symmetry breaking needs to be too small to explain the $g-2$ anomaly. As we shall see, embedding our simplified scenarios within the 3-3-1 models helps alleviate these tensions. 



Our work is structured as follows: in {\it Section} II, we describe the simplified models; in {\it Section} III, we introduce 3-3-1 models and we present our results for 3-3-1 models that feature an inert scalar triplet and exotic charged leptons; in {\it Section} IV, we explore the connection between $g-2$ and collider searches; in {\it Section} V, we make the link between $g-2$ and lepton flavor violation, and in {\it Section} VI, we draw our conclusions. 

\section{Simplified Models}

Several simplified models have been proposed to explain $g-2$ but many of them are excluded by LEP and LHC data \cite{Freitas:2014jla}. In this section, we will discuss two possible simplified models that invoke the presence of exotic charged leptons and an inert scalar doublet.\\

{\bf Inert scalar}\\

An inert scalar doublet ($\phi$) under $SU(2)_L$ with hypercharge $Y=1$ can be introduced to the Standard Model via,

\begin{equation}
    \mathcal{L} =\lambda_{ab} \bar{L}_{aL} \phi e_{bR} + h.c.
    \label{Eqinert1}
\end{equation}where $L_{aL}$ is the Standard Model lepton doublet, $\lambda_{ab}$ is the Yukawa coupling. Such a scalar doublet is subject to several constraints  \cite{Freitas:2014jla}. The couplings to Standard Model fermions  also give rise a set of model-dependent bounds, which depend on whether this new doublet also interacts with quarks, the strength of its couplings to charged leptons, and if it represents a possible dark matter candidate \cite{Wang:2014sda,Hektor:2015zba,Cherchiglia:2017uwv,Wang:2018hnw,Iguro:2019sly}. The contribution to $g-2$ is found to be \cite{Lindner:2016bgg},

\begin{equation}
\Delta a_{\mu}= \frac{m_\mu^2}{8\pi^2 M_\phi^2}\int^1_0 dx \sum_b \frac{\lambda_{2b}^2 x^2(1-x+\epsilon_b)}{(1-x)(1-\lambda^2 x)+x \epsilon_b \lambda^2} 
\label{eqg2}
\end{equation}where $\epsilon_b=m_b/m_\mu$ and $\lambda=m_\mu/M_\phi$.\\

Such an inert scalar can also induce a sizeable branching ratio $\mu \rightarrow e \gamma$ which is severely constrained by data. From Eq.\eqref{Eqinert1} we get,
\begin{equation}
BR(\mu \rightarrow e\gamma) =\frac{3 (4\pi)^3 \alpha_{em} }{4 G_F^2}  |A^M_{e\mu}|^2 
\label{eqBR}
\end{equation}where $A^M$ is a form factor that accounts for the loop correction and carries information about the couplings and masses with,
\begin{equation}
A^M_{ji}= \frac{1}{(4\pi)^2} \lambda_{fj} \lambda_{fi} I_{f,1}^{++}   
\end{equation}where,
\begin{eqnarray}
&& I_{f,1}^{++}=\int dx\,dy \, dz\, \delta(1-x-y-z) \times \nonumber\\
&& \frac{x\,(y+z \, m_j/m_i) + (1-x)\, m_f/m_i}{-xy\, m_i^2 -x z \, m_j^2 + x\, M_\phi^2 + (1-x)\, m_f^2}.    
\end{eqnarray}

This general result can be used in all scenarios explored in this work. \\

{\bf Heavy Charged Leptons}\\

Heavy charged leptons ($E$) represent an interesting class of models to explain the $g-2$ anomaly \cite{Queiroz:2014zfa}. Anomaly cancellation requires the introduction of a vector-like lepton with mass arising from a term like $m_E \bar{E_L} E_R$. If $E$ is a singlet under $SU(2)_L$ it will have interactions of the form,
\begin{equation}
\mathcal{L}= \lambda_{ab} \bar{E}_{aL} \phi \mu_R + h.c
\end{equation}
where $\phi$ is now a scalar singlet under $SU(2)_L$. Another way to generate a correction to $g-2$ via an exotic charged lepton is through the introduction of an inert doublet with,
\begin{equation}
\mathcal{L}= \lambda_{ab} \bar{L}_{aL} \phi E_{bR} + h.c.
\end{equation}
One could also introduce exotic charged leptons ($\psi$) which are doublets under $SU(2)_L$ with,
\begin{equation}
\mathcal{L}= \lambda_{ab} \bar{\psi}_{aL} \phi \mu_{bR} + h.c.
\end{equation}
As stated previously, these exotic charged leptons are required to be vector-like to ensure gauge anomaly cancellation. The contributions to $g-2$ are determined by Eq.\eqref{eqg2} and arise from an interaction between the muon and a scalar, and an exotic charged lepton, if present.  The key differences between these models are the phenomenological constraints they are subject to. While there are important bounds on exotic charged leptons \cite{Altmannshofer:2013zba}, these bounds will turn out to be less stringent than the collider bounds we will consider in this work.

We have reviewed possible simplified models that could accommodate the $g-2$ anomaly through the presence of an inert scalar and exotic charged leptons. We now outline how these simplified models can be embedded in 3-3-1 models.

\section{3-3-1 Models}

3-3-1 models are based on the $SU(3)_C \times SU(3)_L \times \times U(1)_N$ gauge symmetry and were originally introduced since they offer a plausible answer to the number of generations in the Standard Model. These models have been extensively studied in contexts such as dark matter \cite{Fregolente:2002nx,Hoang:2003vj,deS.Pires:2007gi,Mizukoshi:2010ky,Profumo:2013sca,Dong:2013ioa,Dong:2013wca,Cogollo:2014jia,Dong:2014wsa,Dong:2014esa,Carvajal:2017gjj,Montero:2017yvy,Huong:2019vej}, flavor physics \cite{Cabarcas:2012uf,Cabarcas:2012uf,Santos:2017jbv,Barreto:2017xix,Wei:2017ago,Hue:2017lak}, neutrino masses \cite{Cogollo:2010jw,Cogollo:2008zc,Okada:2015bxa,Vien:2018otl,carcamoHernandez:2018iel,Nguyen:2018rlb,Pires:2018kaj,CarcamoHernandez:2019iwh,CarcamoHernandez:2019vih,CarcamoHernandez:2020pnh}, and collider physics \cite{Meirose:2011cs,Coutinho:2013lta,Nepomuceno:2016jyr,Nepomuceno:2019eaz}.

Since $SU(2)_L$ is promoted to $SU(3)_L$, the fermion generations are arranged in the fundamental (or antifundamental) representation of $SU(3)_L$. After the $SU(3)_L \times U(1)_N$ symmetry is spontaneously broken via the vacuum expectation value of a scalar triplet, a remnant $SU(2)_L \times U(1)_Y$ arises \cite{Borges:2016nne}.  The presence of an $SU(3)_L$ triplet implies the existence of new leptons and quarks that contribute to the new non-trivial gauge anomalies. The electric charge operator that preserves the vacuum is found to be,
\begin{equation}
\frac{Q}{e}=\frac{1}{2} (\lambda_3+ \alpha \lambda_8)+ N I = 
\begin{pmatrix}
     1/2 (1+\frac{\alpha}{\sqrt{3}}) +N  \\
    1/2 (-1+ \frac{\alpha}{\sqrt{3}}) +N \\
     - \frac{ \alpha}{\sqrt{3}} +N
\end{pmatrix},
\end{equation}
where $\lambda_{3},\lambda_8$ and $I$ are the generators of $SU(3)_L$ and $U(1)_N$, respectively. The parameters $\alpha$ and N are in principle free parameters. 

Nevertheless, as we need to recover the Standard Model spectrum, the first two components of the triplet should be a neutrino and a charged lepton. Therefore, $\alpha / \sqrt{3}=-(2N+1)$. The third component of the triplet should have a $U(1)_N$ quantum number equal to $3N+1$.  If one takes  $N=0$, then the third component of the triplet would be a positively charged field. The Minimal 3-3-1 model \cite{Pisano:1991ee} and the 3-3-1 model with exotic charged leptons \cite{Ponce:2001jn,Ponce:2002fv,Anderson:2005ab,Cabarcas:2013jba} are based on this choice.  However, if one instead chooses $N=-1/3$ then the third field component of the lepton triplet would be either a right-handed neutrino ($\nu_R^c$) or a heavy lepton ($N$). These possibilities are known as 3-3-1 models with right-handed neutrinos \cite{Hoang:1996gi,Hoang:1995vq} and 3-3-1 models  with a left-handed neutral lepton  \cite{Mizukoshi:2010ky,Catano:2012kw}.  The popular Economical 3-3-1 model is essentially the 3-3-1 model with right-handed neutrinos where two scalar triplets form the scalar sector \cite{Dong:2006mg,Dong:2008ya,Berenstein:2008xg,Martinez:2014lta}. 

It has been recently shown that none of these models are capable of accommodating the $g-2$ anomaly due to the existence of stringent collider bounds (see \cite{Ky:2000ku,Kelso:2013zfa,Binh:2015cba,Cogollo:2017foz,DeConto:2016ith} for $g-2$ studies in the context of 3-3-1 models). The masses of the gauge bosons are proportional to the energy scale at which $SU(3)_L \times U(1)_N \rightarrow SU(2)_L \times U(1)_Y$. Since the LHC places strong lower bounds on the masses of the gauge bosons, one can convert these bounds into limits on the energy scale of the 3-3-1 symmetry breaking. A common feature among all these models is that the energy scale of 3-3-1 symmetry breaking needs to be too low ($\sim 1$~TeV) to explain $g-2$. This is forbidden by collider searches.  In summary, the most popular 3-3-1 models in the literature cannot address $g-2$. 


We now discuss the embedding of the simplified Lagrangians presented previously within the 3-3-1 model. We will take the 3-3-1 model with neutral heavy leptons as a benchmark, but our results can be easily extended to all five models mentioned earlier. In the 3-3-1 model with heavy neutral leptons, the leptons are arranged as follows,
\begin{equation} f^{a}_L =
\begin{pmatrix}
   \nu^a  \\
    l^a  \\
    N^a
\end{pmatrix} ;\,\, l^a_R, N^a_R.
\label{lrr1}
\end{equation}
where $a$ runs from one to three and $N^a$ are heavy neutral leptons. We do not consider the hadronic sector since it is not relevant to our discussion. The fermion masses are generated by  spontaneous symmetry breaking governed by the three scalar triplets:
\begin{equation}
    \chi=\begin{pmatrix}
         \chi^0\\
         \chi^{-}\\
         \chi^{0\prime}
    \end{pmatrix},
         \rho=\begin{pmatrix}
         \rho^+\\
         \rho^0\\
         \rho^{+\prime}
    \end{pmatrix},
    \eta=\begin{pmatrix}
         \eta^0\\
         \eta^-\\
         \eta^{0\prime}
    \end{pmatrix}.
    \label{tripletscalars2}
\end{equation}
One could also successfully generate the fermion masses with two scalar triplets as occurs in the Economical 3-3-1 model \cite{Dong:2006mg,Dong:2008ya}. We stress that the details are not relevant because our results can still be applied to this model. The leptons acquire mass via the Yukawa term in the Lagrangian,
\begin{equation}
\mathcal{L} \supset G_{ab} \bar{f}_{aL} \rho e_{bR} + g^\prime_{ab} \bar{f}_{aL} \chi N_{bR} + h.c.
\end{equation}
When the neutral fields $\chi^{0\prime}$ and $\rho^0$ develop a non-zero vacuum expectation value, the leptons acquire a mass term. Here we label $\langle \chi^{0\prime}\rangle = v_{\chi}$ and  $\langle\rho^0, \eta^0 \rangle = v_\eta=v_\rho \equiv 246/\sqrt{2}$. The neutral and vector currents involving new gauge bosons are found to be \cite{Hoang:1995vq,Hoang:1996gi},
\begin{equation}
{\cal L}^{NC} \supset
\bar{f}\, \gamma^{\mu} [g_{V}(f) + g_{A}(f)\gamma_5]\, f\,
Z'_{\mu}, \label{eqNC}
\end{equation}  with
\be
g_{V}(f) = \frac{g}{4 c_W} \frac{(1 -
4 s_W^2)}{\sqrt{3-4s_W^2}},\
g_{A}(f) = -\frac{g}{4 c_W \sqrt{3-4s_W^2}},\nonumber
\label{gvga}
\ee and,
\bea
{\cal L^{CC}} \supset - \frac{g}{\sqrt{2}}\left[
\overline{N_L}\, \gamma^\mu \bar{l_L}\, W^{\prime -}_\mu  \right].
\label{eqCC}
\eea where,
\bea
M_{Z^{\prime}}^2 = \frac{g^2}{4(3-4s_W^2)}\left( 4 c_W^2 v_{\chi}^2 + \frac{v_{\rho}^2}{c_W^2}
+ \frac{v_{\eta}^2(1-2s_W^2)^2}{c_W^2} \right),
\label{Zprimemass}\nonumber\\
\eea and
\begin{equation}
M_{W^{\prime}}^2 =M_{X^0}^2 = \frac{g^2}{4}\left( v_{\eta}^2 + v_{\chi}^2 \right).
\label{wprimemass}
\end{equation}

The largest corrections to $g-2$ come from interactions involving new gauge bosons. It has been shown that the most popular 3-3-1 models are incapable of addressing $g-2$ due to the existence of collider bounds as we describe below.

\subsection{Collider Bounds}

The most important bounds on 3-3-1 models come  from LHC searches for new $Z^\prime$ gauge bosons. These limits rely on dilepton resonance searches. Assuming that the $Z^\prime$ decays only into charged leptons the lower mass bound imposes $M_{Z^\prime} > 4$~TeV \cite{Lindner:2016bgg}, which translates into $v_{\chi}>12$~TeV using Eq.\ref{Zprimemass}. For easy comparison between $M_{Z^\prime}$ and $v_{\chi}$, we note that $M_{Z^\prime}=0.395 v_{\chi}$. 

We note that the heavy neutral lepton may be sufficiently light and become kinematically accessible for decays for the $Z^\prime$ boson. The 3-3-1 models typically feature exotic quarks which can also be lighter than $M_{Z^\prime}/2$. Assuming that all exotic quarks and heavy neutral leptons represent kinematically accessible decay channels, the LHC bounds significantly weaken. Our implementation of the model takes the branching ratio into charged leptons to be between 50\%-60\% in agreement with \cite{deJesus:2020ngn}. The bounds derived from dilepton resonance searches are proportional to the branching ratio into charged leptons. Therefore, we can conservatively state that in light of these new decay modes a conservative LHC bound should read,
\begin{equation}
{\rm LHC\, 13~TeV}: \,\,   M_{Z^\prime} > 2\, {\rm TeV}; \,\, v_{\chi}> 5\, {\rm TeV}.
\label{eqLHC}
\end{equation}
As stated in the Introduction, we have assumed the conservative PDG $\Delta a_\mu$ value and for this reason, we will also assume this conservative LHC bound. We stress that our overall conclusions will not change if one decides to alter the LHC bound ($M_{Z^\prime} > 4$~TeV) because there is enough freedom to change other parameters and still obtain qualitatively similar results in regard to $g-2$.


\begin{figure}
    \centering
    \includegraphics[width=0.95\columnwidth]{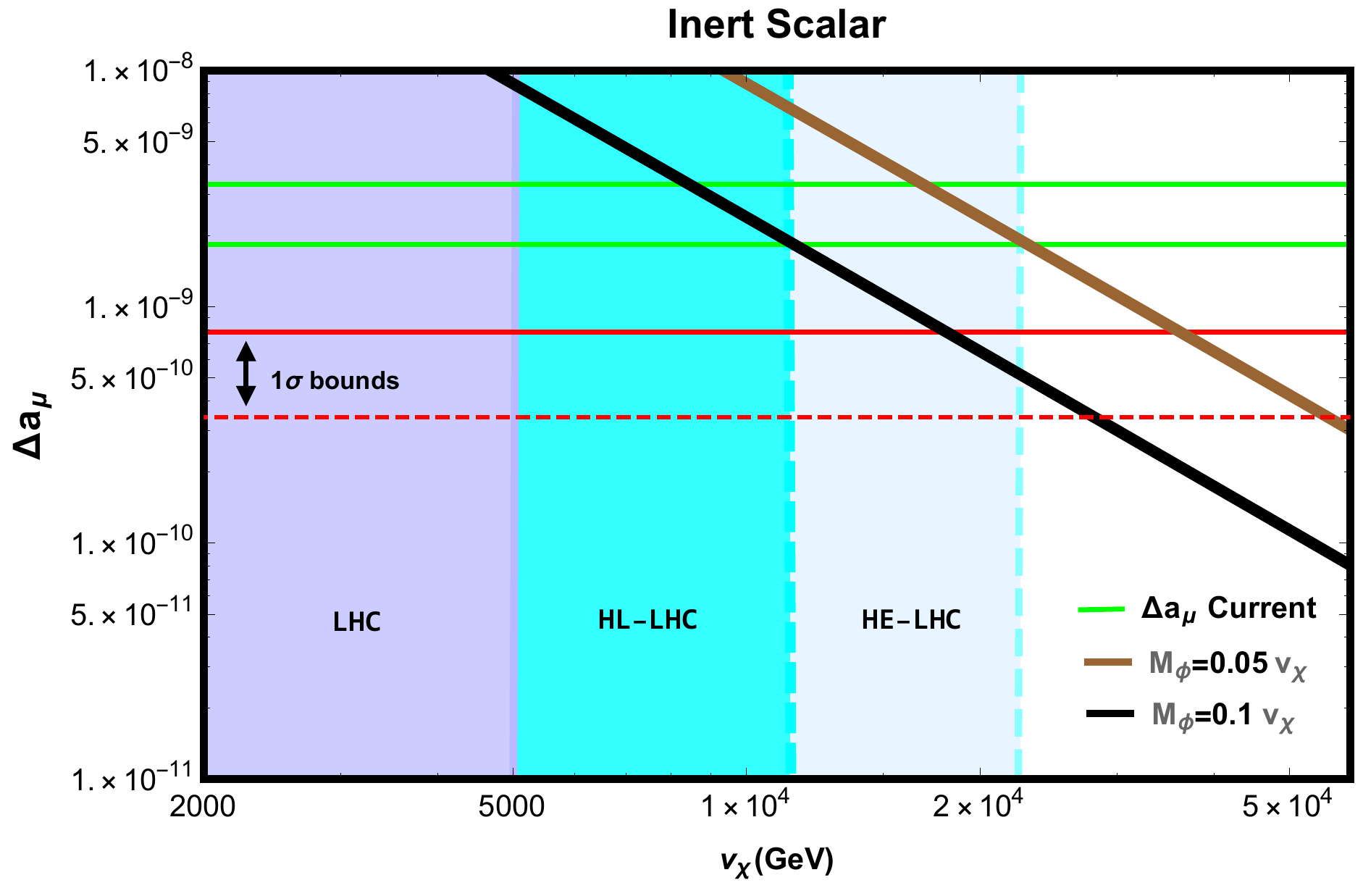}
    \caption{Overall correction to $g-2$ as a function of the energy scale of 3-3-1 symmetry breaking for $M_{\phi} = 0.1 v_{\chi}$ (black curve) and $M_{\phi}=0.05 v_{\chi}$ (brown curve) due the presence of an inert scalar triplet. In this plot, we considered $\lambda_{22} = 1$. The green lines delimit the region of parameter space that explains $g-2$. The LHC, HL-LHC and HE-LHC limits are displayed according to Eqs.\eqref{eqLHC}-\eqref{eqHLLHC}. The horizontal red lines represent the current and projected $1\sigma$ bounds by requiring the new physics contribution to be within the $1\sigma$ error bar. One can clearly see that if we take $v_{\chi}\sim 10-12$~TeV we can  explain the $g-2$ anomaly while being consistent with current LHC bounds. See text for further details. }
    \label{fig:inert}
\end{figure}

We now go on to discuss projected constraints coming from the High-Luminosity (HL-LHC) and High-Energy LHC (HE-LHC) on these scenarios. HL-LHC refers to LHC configuration at $14$~TeV center-of-mass energy with $3 \rm{ab}^{-1}$ of integrated luminosity. The HE-LHC denotes a $27$~TeV center-of-mass energy with $15 \rm{ab}^{-1}$ of integrated luminosity. For dilepton resonance searches the number of signal and background events scale equally with energy and luminosity. Thus, the bounds  on the number of signal events at each value of the $Z^\prime$ mass can be obtained in terms of the number of background events. We note that there are implicit assumptions about the event acceptance and efficiency rate, which are assumed to be independent of the dilepton invariant mass. It has been shown that these assumptions are reasonable for resonance searches under the narrow width approximation \cite{CidVidal:2018eel} and as long as null results are reported. Therefore, one can use the code described in \cite{Thamm:2015zwa} to determine the HL-LHC and HE-LHC reach and conclude that
\begin{eqnarray}
{\rm HL-LHC}:\,\, M_{Z^\prime} > 4.5\, {\rm TeV};\,\, v_{\chi}>11.4 \, {\rm TeV},\\
{\rm HE-LHC}:\,\, M_{Z^\prime} > 8.9\, {\rm TeV};\,\, v_{\chi}>22.5\, {\rm TeV}.
\label{eqHLLHC}
\end{eqnarray}
\begin{figure}
    \centering
    \includegraphics[width=0.95\columnwidth]{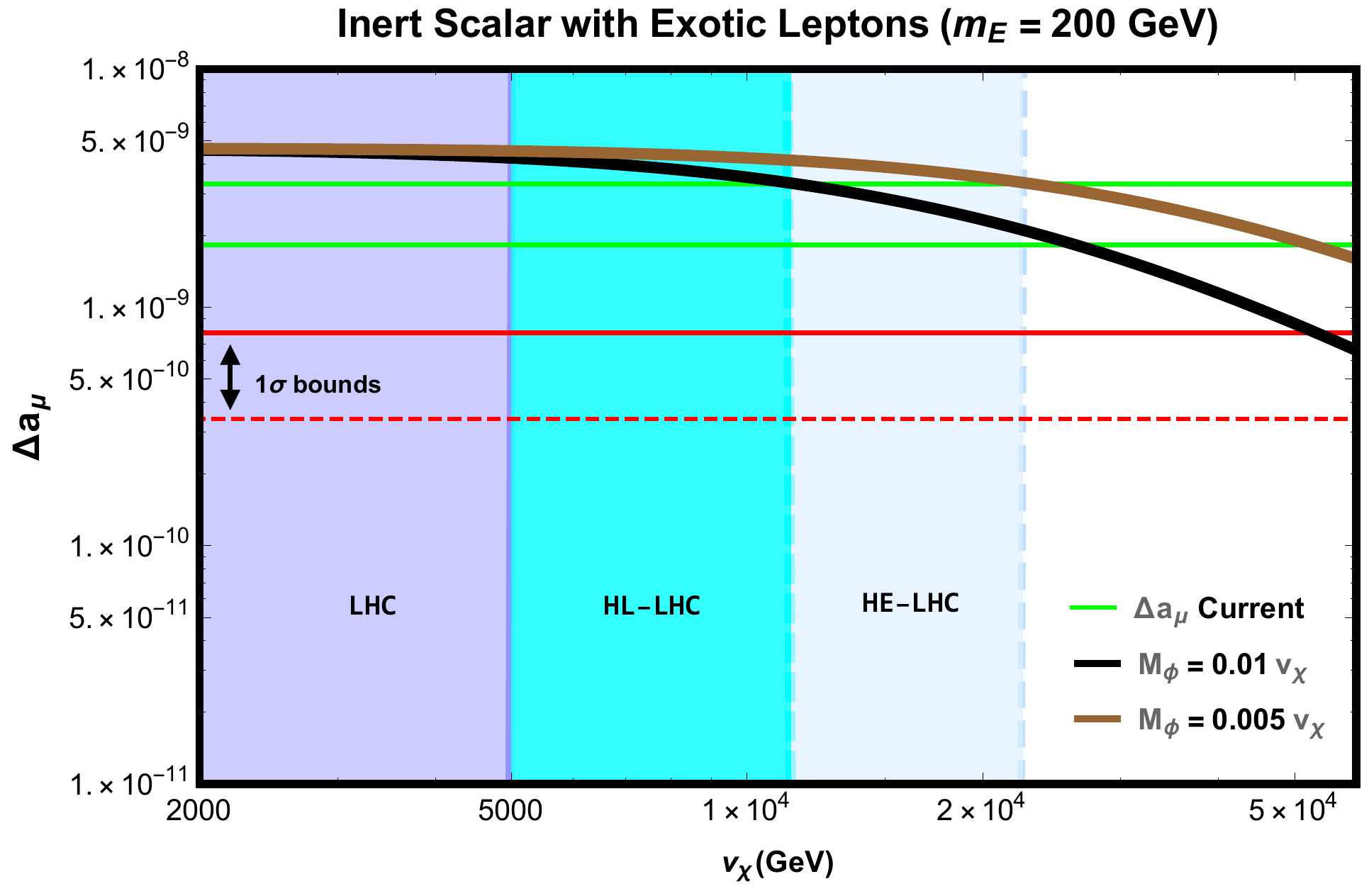}
    \caption{Overall correction to $g-2$ as a function of the energy scale of 3-3-1 symmetry breaking for $M_{\phi} = 0.01v_{\chi}$ (black curve) and $M_{\phi}= 0.005 v_{\chi}$ (brown curve) due the presence of an exotic lepton with $m_E=200$~GeV, for the case where $\lambda_{22} = 2$. The green lines delimit the region of parameter space that explains $g-2$. The LHC, HL-LHC and HE-LHC bounds are displayed according to Eqs.\eqref{eqLHC}-\eqref{eqHLLHC}. The horizontal red lines represent the current and projected $1\sigma$ bounds by requiring the new physics contribution to be within the $1\sigma$ error bar. One can clearly see that if we take $v_{\chi}\sim 20$~TeV we explain the $g-2$ anomaly while being consistent with current and future collider bounds. See the text for further details.}
    \label{figmE1}
\end{figure}

\begin{figure}
    \centering
    \includegraphics[width=0.95\columnwidth]{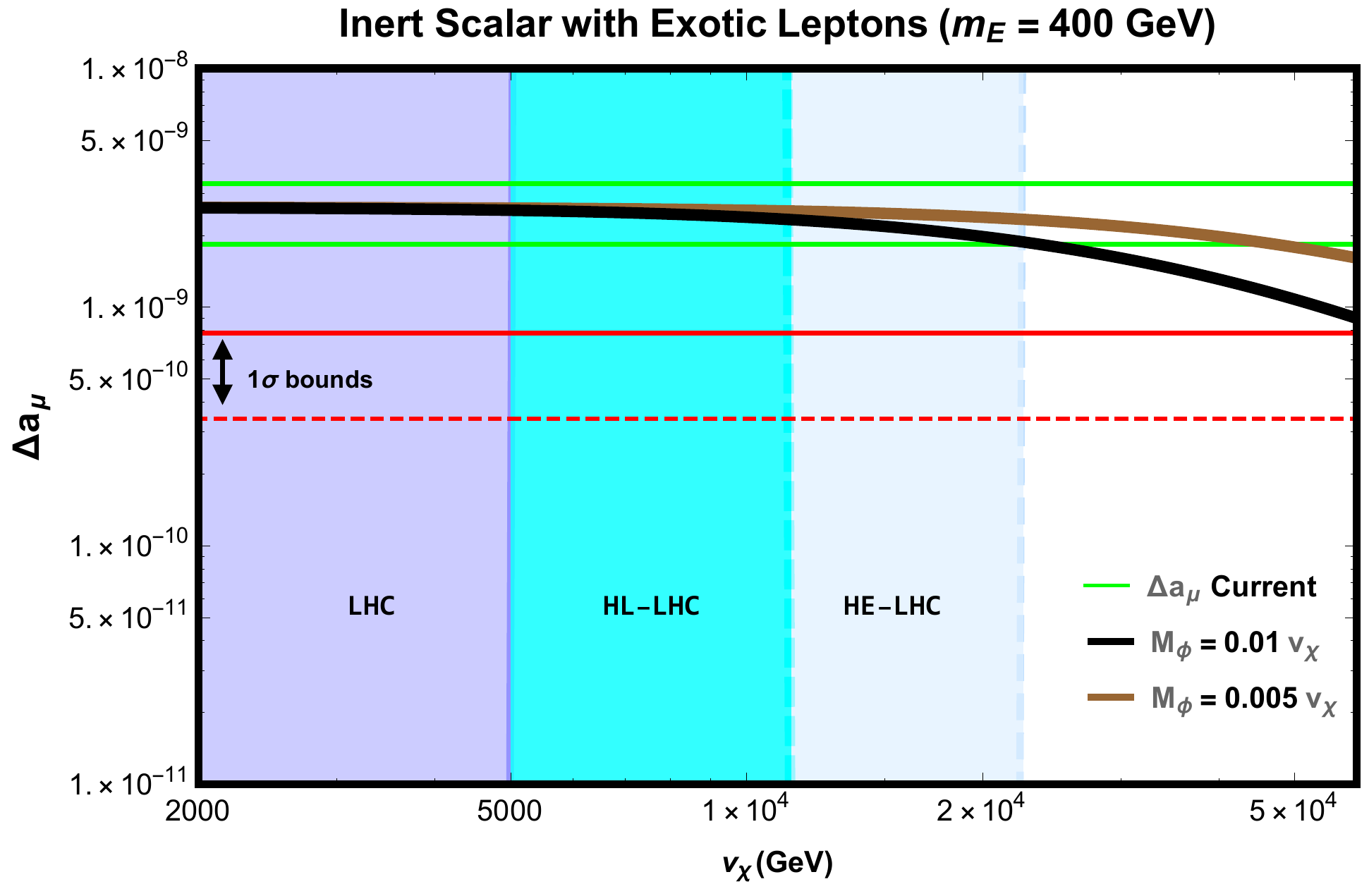}
    \caption{Same as FIG.\ref{figmE1} but for $m_E=400$~GeV and $\lambda_{22} = 3$. Now the scale of symmetry breaking is in agreement to higher values, until approximately $25$~TeV for $M_\phi \sim 0.01v_\chi$, and $40$~TeV for $M_\phi \sim 0.005 v_\chi$. Both scenarios can be probed by HE-LHC. For $M_\phi \sim 0.01v_\chi$, the parameter space is completely probed by HE-LHC. It is interesting to see that the current and projected $g-2$ bounds are stronger than those arising from collider searches.}
    \label{figmE2}
\end{figure}

\section{Muon Anomalous Magnetic Moment and Collider Physics}

In this Section, we embed the simplified scenarios involving  an inert scalar and exotic leptons introduced earlier into the 3-3-1 models. Our purpose is to study the $g-2$ anomaly and correlate that with potential positive signals in $\mu \rightarrow e\gamma$ decay, in agreement with existing and future collider bounds.\\

{\bf Inert Scalar Triplet}\\

Since the relevant symmetry is  $SU(3)_L$, an inert scalar triplet can be trivially added.  Its contribution to $g-2$ is proportional to the  mass of the muon, which would imply a suppressed correction to $g-2$. Thus, we can add
\begin{equation}
    \mathcal{L}= \lambda_{ab} \bar{f}_{aL} \phi e_{bR}.
\end{equation}
The inert scalar triplet acquires a mass term from the quartic scalar potential term that goes as $\lambda \chi^\dagger \chi \phi \phi$. Therefore, $M_{\phi} \sim \lambda v_{\chi}$. 

We display our results for $M_{\phi} =(0.05,0.1) v_{\chi}$. We exhibit the correction to $g-2$ in FIG.\ref{fig:inert} for $M_{\phi} = 0.1 v_{\chi}$ (black curve) and $M_{\phi}=0.05 v_{\chi}$ (brown curve) and overlay the current and projected collider bounds. In this plot, we considered the Yukawa coupling of the muon with the scalar triplet as $\lambda_{22}=1$, obeying the limits from unitarity, which implies that $\lambda_{22}<4\pi$. The green lines delimit the region of parameter space that explains $g-2$. The LHC, HL-LHC and HE-LHC limits are superimposed according to Eqs.\eqref{eqLHC}-\eqref{eqHLLHC}. The horizontal red lines represent the $g-2$ bounds by requiring  the correction to be within the current ($78\times 10^{-11}$) and projected ($34\times 10^{-11}$) $1\sigma$ error. From FIG.\ref{fig:inert} we conclude that a scale of symmetry breaking around $10-15$~TeV is necessary to accommodate $g-2$. It is important to stress that our choices for $\lambda=0.05$ and $0.1$ are not random. We found that $\lambda$ should be around $0.1$ to be able to explain $g-2$ while still resulting in an energy scale accessible by future colliders. Other values for $\lambda$ are possible but then this complementarity with collider searches is lost. \\ 

\begin{figure}
    \centering
    \includegraphics[width=\columnwidth]{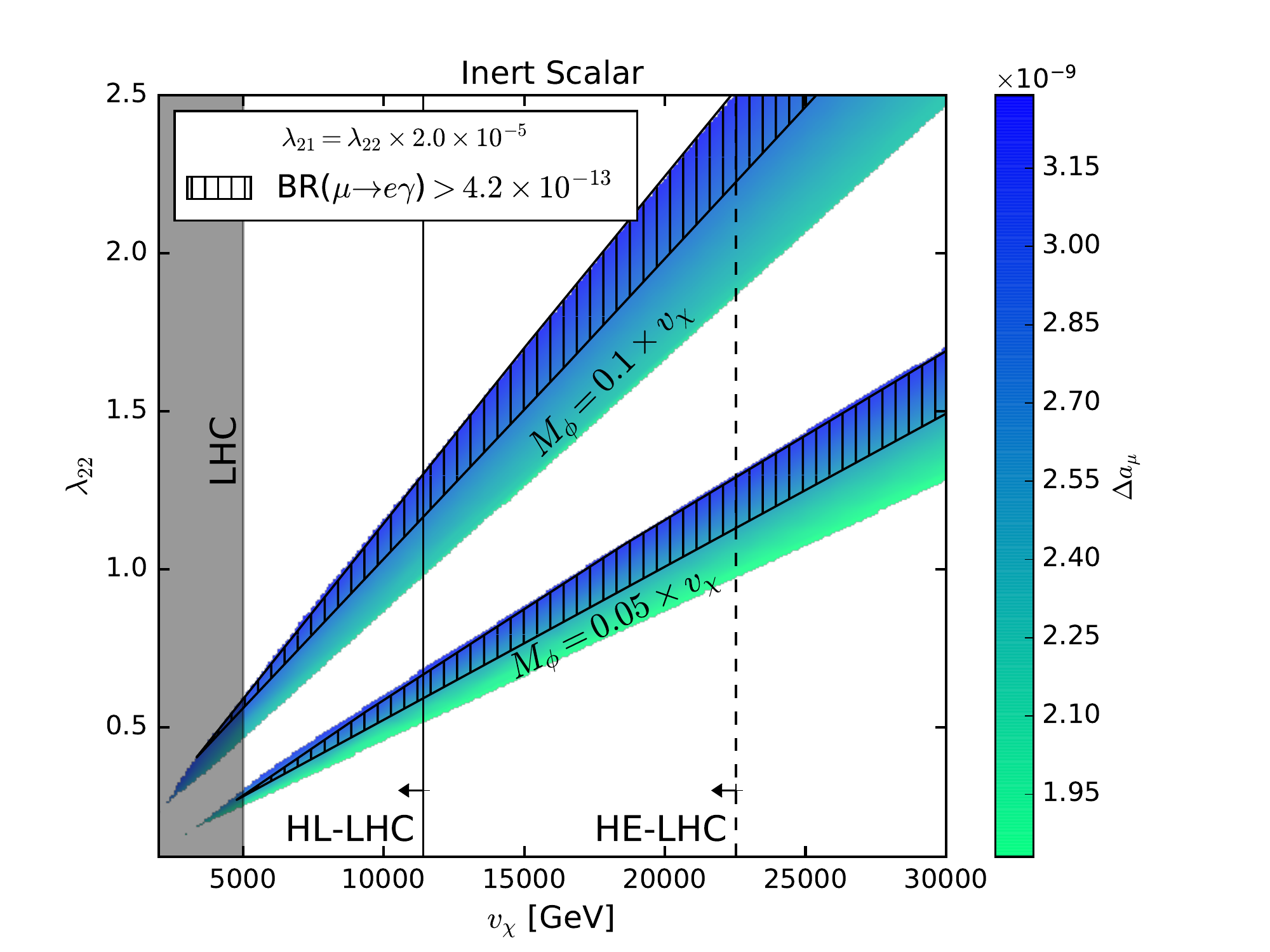}
    \caption{Contour plot of the regions that explain $g-2$ as a function of the scale of symmetry breaking, $v_{\chi}$. We assumed $\lambda_{22}= 2 \lambda_{21} \times 10^4$. The two distinct regions are for $M_{\phi}= 0.05v_{\chi}$ and $M_{\phi}= 0.1v_{\chi}$. The vertical blue lines delimit the HL-LHC and HE-LHC projected exclusion regions.}
    \label{figCla1}
\end{figure}

{\bf Exotic Charged Lepton}\\

We now explore the addition of vector-like charged leptons. Similar to the previous case, we can introduce a Lagrangian term of the type,
\begin{equation}
    \mathcal{L}= \lambda_{ab} \bar{E}_{aL} \phi \mu_{bR}.
\end{equation}
where $\phi$ is singlet scalar field and $E_{a}$ are vector-like charged leptons, which we assume to be mass degenerate for simplicity. The mass of the vector-like lepton arises from a bare mass term as described in {\it Section} II. The key difference between this scenario and the previous one is the presence of an additional free parameter, which is the mass of the vector-like lepton, $m_E$. The mass of this scalar singlet will also be proportional to $v_{\chi}$, i.e $M_\phi \sim \lambda^\prime v_\chi$, where $\lambda^\prime$ is the dimensionless quartic coupling in the scalar potential, similar to the previous case. We exhibit our findings for different values of $m_E$ in FIGs.\ref{figmE1}-\ref{figmE2}. A solution to $g-2$ via the introduction of a scalar singlet was also explored in  \cite{CarcamoHernandez:2020pxw}, but in a different context. 

From FIGs.\ref{figmE1}-\ref{figmE2}, we conclude that $v_{\chi}\sim 20$~TeV and $v_{\chi}\sim 40$~TeV are favored for $M_{\phi}=0.005 v_{\chi}$ and $M_{\phi}= 0.01 v_{\chi}$, respectively, where we considered the Yukawa couplings to the scalar singlet to be $\lambda_{22}=2.0$ for the $m_E=200$ GeV case. Either way, it is interesting to see that there is a strong complementarity between future colliders and $g-2$. Again notice that $g-2$ gives rise to stronger bounds than colliders by requiring the new physics corrections to be below the $1\sigma$ error bar. 
\begin{figure}
    \centering
    \includegraphics[width=1.\columnwidth]{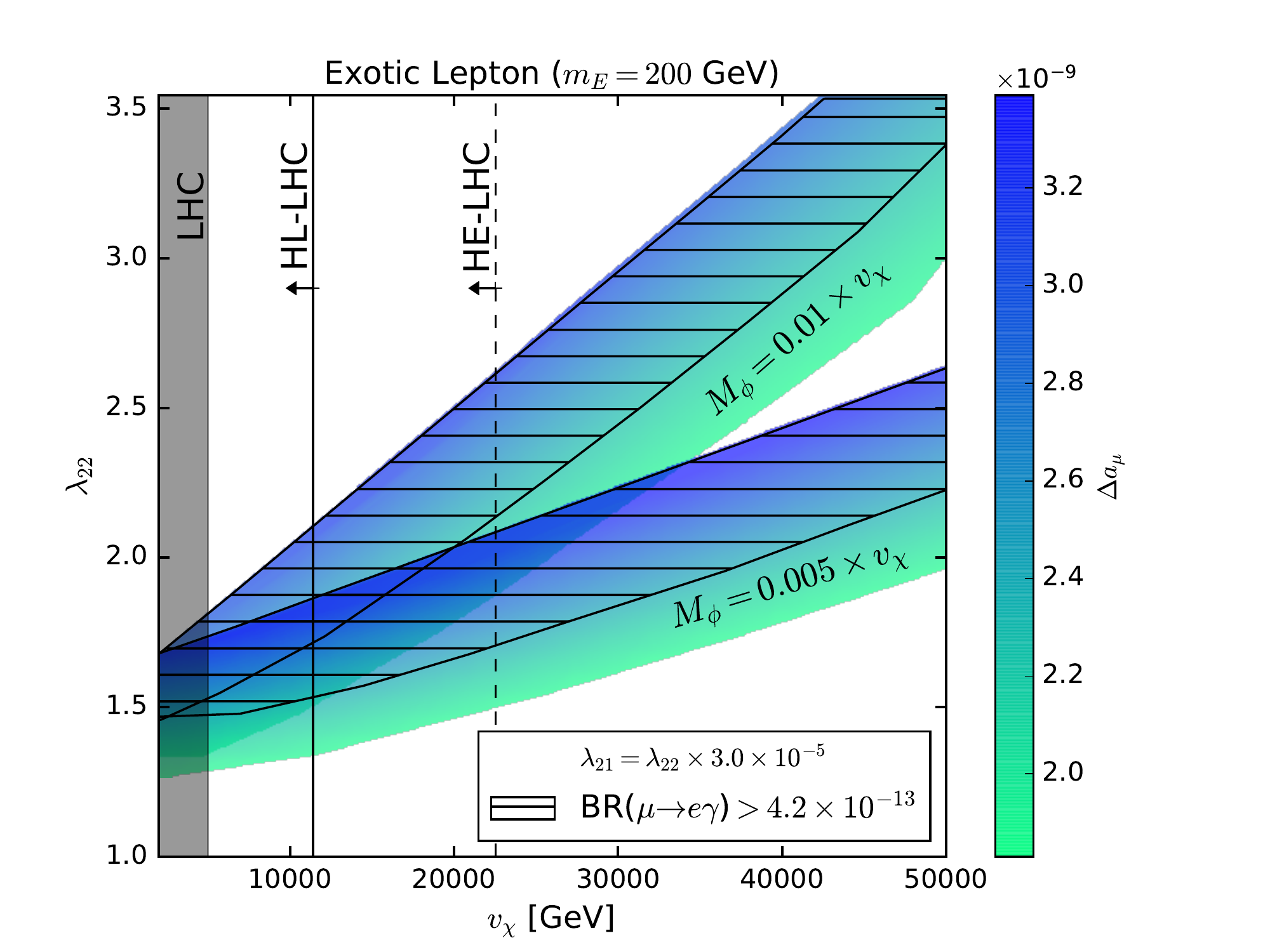}
    \caption{Contour plot of the regions that explain $g-2$ as a function of the scale of symmetry breaking in the presence of an exotic charged lepton with $m_E=200$~GeV. We assumed $\lambda_{21}= 3.0 \lambda_{21} \times 10^{-5}$. The two distinct regions are for $M_{\phi}= 0.005v_{\chi}$ and $M_{\phi}= 0.01v_{\chi}$. The shaded black region is excluded by LHC searches and the vertical black lines delimit the HL-LHC (continuous) and HE-LHC (dashed) projected exclusion regions.}
    \label{figCla2}
\end{figure}

\begin{figure}
    \centering
    \includegraphics[width=1.\columnwidth]{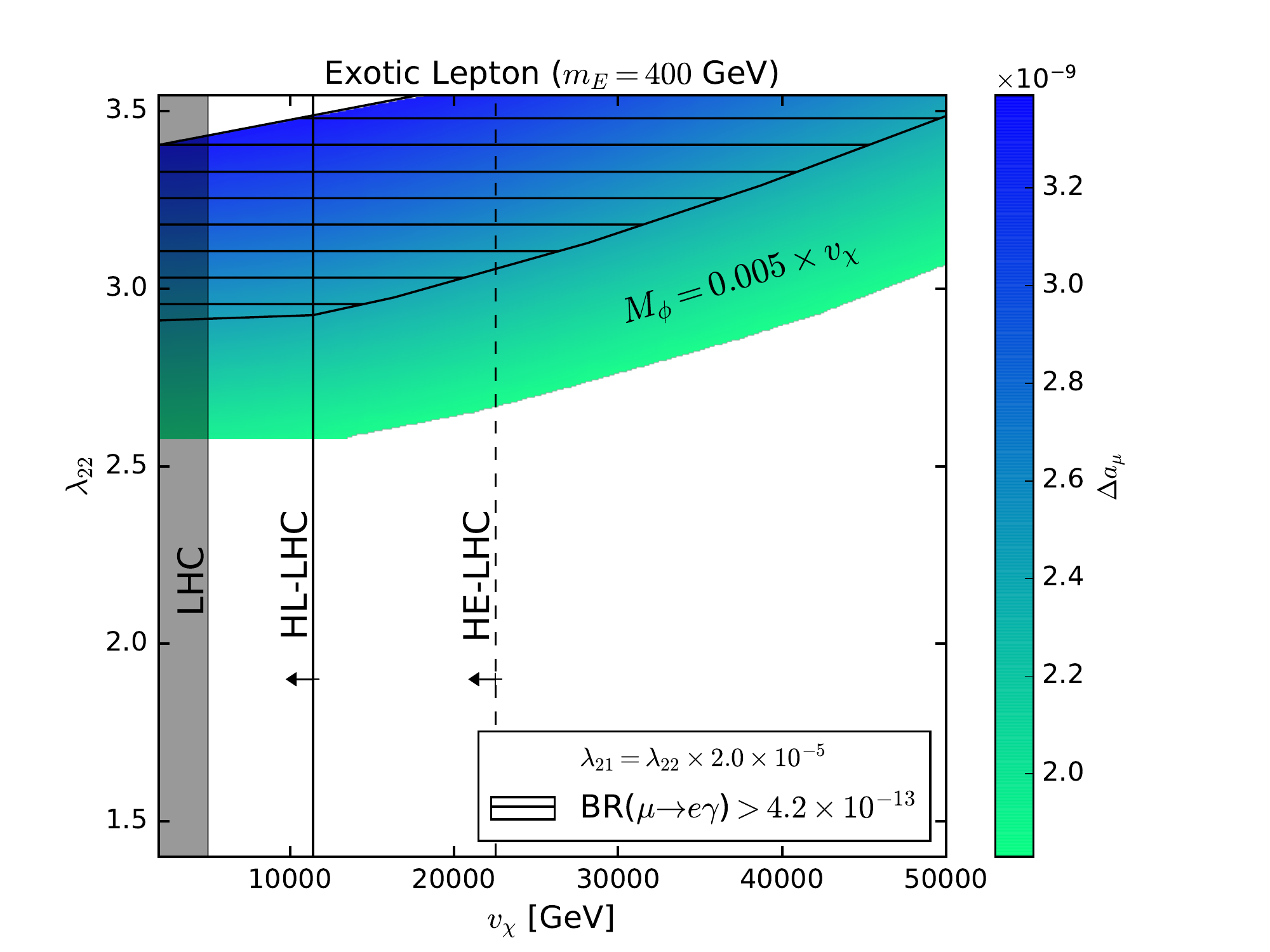}
    \caption{Same as FIG.\ref{figCla2} but for $m_E=400$~GeV and $M_{\phi}= 0.005v_{\chi}$. }
    \label{figCla3}
\end{figure}

For $m_E=400$~GeV, the situation changes a bit as the scale of symmetry breaking explains  $g-2$ up to higher values ($2\, {\rm TeV} < v_{\chi} < 40\, {\rm TeV}$), leaving an unexplored region above $v_\chi>23$~TeV to future colliders when we choose $M_{\phi}= 0.005v_{\chi}$, and completely probed when we take $M_{\phi}= 0.01v_{\chi}$, both for $\lambda_{22}=3.0$.

\section{Connection with Lepton Flavor Violation}

The interplay between $g-2$ and collider bounds has been explored thus far. We now go on to a discussion of the complementarity of our results with lepton flavor violation.

 To correlate our findings with lepton flavor violation, we assume benchmark values for the off-diagonal Yukawa couplings. Our results are based on Eq.\eqref{eqBR}. In FIG.\ref{figCla1}, we show a contour plot for the inert scalar model in the $\lambda_{22}$-$v_{\chi}$ plane with the regions that address $g-2$ for $M_{\phi}=0.05 v_{\chi}$ and $M_{\phi}=0.1 v_{\chi}$ and superimpose the bounds on lepton flavor violation arising from future colliders. The hashed regions are excluded by $\mu \rightarrow e\gamma$ and the black vertical lines delimit the HL-LHC (continuous) and HE-LHC (dashed) projected limits according to Eq.\eqref{eqHLLHC}. To exhibit all these observables in the same plane we assumed $\lambda_{21}= 2\, \lambda_{22} \times 10^{-5}$. Notice that the diagonal coupling, $\lambda_{22}$, should be much larger than the off-diagonal coupling, $\lambda_{21}$, otherwise the bound coming from $\mu \rightarrow e\gamma$ would rule out the entire region of parameter space in FIG.\ref{figCla1}. Notice that if we take $v_{\chi} \sim 12$~TeV, we can explain $g-2$ and leave signatures at the HE-LHC and MEG II detectors. For any value chosen for the scale of symmetry breaking $v_{\chi}$, we find similar conclusions. This is true because we chose a value for the ratio $\lambda_{21}/\lambda_{22}$, to be able to show regions that are either excluded or consistent with $\mu \rightarrow e\gamma$. If we had adopted larger values for $\lambda_{21}$ then the $\mu \rightarrow e\gamma$ decay would have excluded the entire viable region for $g-2$. We could draw a similar conclusion for $M_{\phi}= 0.1 v_{\chi}$. The relation between $M_{\phi}$ and $v_{\chi}$ was  obtained by requiring that the parameter that resolves $g-2$ is testable at future collider and lepton flavor violation searches. Moreover, we can also conclude that $\mu \rightarrow e\gamma$ exclusion region extends the HE-LHC one. Again, this demonstrates the importance of searching for new physics with complementarity between different observables and detectors.

In the presence of a vector-like exotic charged lepton the situation changes, as we have shown before. Now the connection to lepton flavor violation can be observed as well. In FIGs.\ref{figCla2} and \ref{figCla3}, we display contour plots in the region of parameter space that accommodates the $g-2$ anomaly and overlay that with projected collider and lepton flavor violation constraints. The only difference between these figures is the value adopted for the exotic charged lepton masses that goes from $m_E=200$~GeV in  FIG.\ref{figCla2}  to $m_E=400$~GeV in FIG.\ref{figCla3}. Here we took $\lambda_{21}=3\, \lambda_{22} \times 10^{-5}$, but now we explored the scenarios with $M_{\phi}=0.005 v_{\chi}$ and $M_{\phi}=0.01v_{\chi}$. We stress that these relations between $M_{\phi}$ and $v_{\chi}$ are obtained by requiring a strong degree of complementarity between $g-2$, lepton flavor violation and collider physics. Focusing on Fig.\ref{figCla2}, we can easily conclude that the region with $12 {\rm TeV} < v_{\chi} < 22 {\rm TeV}$ can be tested by future HL-LHC and HE-LHC colliders and $\mu \rightarrow e\gamma$ decay searches. This conclusion holds true regardless of the exotic charged lepton mass. The quantitative predictions for $g-2$ and $BR(\mu \rightarrow e\gamma)$ obviously change with the choice of the exotic lepton mass, as one can clearly see from FIG.\ref{figCla3}.


The scale of symmetry breaking that addresses $g-2$ and allows for a signal in $\mu \rightarrow e\gamma$ is tied to the masses of the neutral ($Z^\prime$) and charged gauge bosons $(W^{\prime \pm})$ . Therefore, if we observed positive signals in $g-2$ and $\mu \rightarrow e\gamma$, we would be able to predict the masses of the new gauge bosons to be observed at future LHC searches for dilepton resonances and charged lepton plus missing energy events which are the golden channel to search for these gauge bosons. This correlation could thus be used to discriminate our model from others in the literature. 


%
\section{Conclusions}

We presented  simplified scenarios that can accommodate the muon $g-2$ anomaly, and discussed their embedding into 3-3-1 models. In the augmented 3-3-1 models with charged vector-like leptons, we were able to accommodate $g-2$ while also being in agreement with collider bounds on the $W^{\prime}$ and $Z^{\prime}$ bosons that appear in these models.  This conclusion is valid for other avatars of the 3-3-1 model as well. Obviously, the quantitative results depend on the values assumed for the yukawa couplings and masses of the vector-like fermions, and we explored the possibility of non-zero off-diagonal couplings to exploit the complementarity between $g-2$, collider searches and lepton flavor violation. If the yukawa couplings are taken to be suppressed the complementarity between different searches is lost. The confluence of various experimental results, especially those coming from the rare muon decay $\mu \to e\gamma$, turned out to be quite restrictive, as can be seen from FIG. 4, FIG. 5, and FIG.6.  More optimistically, the solutions investigated here may explain $g-2$ anomaly and induce signals at upcoming searches for the $\mu \rightarrow e\gamma$ rare decay and future colliders such as the HL-LHC and HE-LHC.


\section*{Acknowledgement}

SK acknowledges support from CONICYT-Chile Fondecyt No. 1190845 and 
ANID-Chile PIA/APOYO AFB180002. ASJ acknowledges support from CAPES. CS is supported by MEC (Ministério da Educação) and UFRN. FSQ thanks CNPq grants 303817/2018-6 and 421952/2018-0, and ICTP-SAIFR FAPESP grant 2016/01343-7 for the financial support. FSQ thanks Andres Bello and UFRGS for the hospitality where part of this was partly done. This work was supported by the
Serrapilheira Institute (grant number Serra-1912-31613). KS is supported by DOE Grant DE-SC0009956.  We thank the High Performance Computing Center (NPAD) at UFRN for providing computational resources.


\bibliography{muon_bbl}

\begin{thebibliography}{95}%
\makeatletter
\providecommand \@ifxundefined [1]{%
 \@ifx{#1\undefined}
}%
\providecommand \@ifnum [1]{%
 \ifnum #1\expandafter \@firstoftwo
 \else \expandafter \@secondoftwo
 \fi
}%
\providecommand \@ifx [1]{%
 \ifx #1\expandafter \@firstoftwo
 \else \expandafter \@secondoftwo
 \fi
}%
\providecommand \natexlab [1]{#1}%
\providecommand \enquote  [1]{``#1''}%
\providecommand \bibnamefont  [1]{#1}%
\providecommand \bibfnamefont [1]{#1}%
\providecommand \citenamefont [1]{#1}%
\providecommand \href@noop [0]{\@secondoftwo}%
\providecommand \href [0]{\begingroup \@sanitize@url \@href}%
\providecommand \@href[1]{\@@startlink{#1}\@@href}%
\providecommand \@@href[1]{\endgroup#1\@@endlink}%
\providecommand \@sanitize@url [0]{\catcode `\\12\catcode `\$12\catcode
  `\&12\catcode `\#12\catcode `\^12\catcode `\_12\catcode `\%12\relax}%
\providecommand \@@startlink[1]{}%
\providecommand \@@endlink[0]{}%
\providecommand \url  [0]{\begingroup\@sanitize@url \@url }%
\providecommand \@url [1]{\endgroup\@href {#1}{\urlprefix }}%
\providecommand \urlprefix  [0]{URL }%
\providecommand \Eprint [0]{\href }%
\providecommand \doibase [0]{http://dx.doi.org/}%
\providecommand \selectlanguage [0]{\@gobble}%
\providecommand \bibinfo  [0]{\@secondoftwo}%
\providecommand \bibfield  [0]{\@secondoftwo}%
\providecommand \translation [1]{[#1]}%
\providecommand \BibitemOpen [0]{}%
\providecommand \bibitemStop [0]{}%
\providecommand \bibitemNoStop [0]{.\EOS\space}%
\providecommand \EOS [0]{\spacefactor3000\relax}%
\providecommand \BibitemShut  [1]{\csname bibitem#1\endcsname}%
\let\auto@bib@innerbib\@empty
\bibitem [{\citenamefont {Bennett}\ \emph {et~al.}(2002)\citenamefont {Bennett}
  \emph {et~al.}}]{Bennett:2002jb}%
  \BibitemOpen
  \bibfield  {author} {\bibinfo {author} {\bibfnamefont {G.~W.}\ \bibnamefont
  {Bennett}} \emph {et~al.} (\bibinfo {collaboration} {Muon g-2}),\ }\href
  {\doibase 10.1103/PhysRevLett.89.129903, 10.1103/PhysRevLett.89.101804}
  {\bibfield  {journal} {\bibinfo  {journal} {Phys. Rev. Lett.}\ }\textbf
  {\bibinfo {volume} {89}},\ \bibinfo {pages} {101804} (\bibinfo {year}
  {2002})},\ \bibinfo {note} {[Erratum: Phys. Rev. Lett.89,129903(2002)]},\
  \Eprint {http://arxiv.org/abs/hep-ex/0208001} {arXiv:hep-ex/0208001 [hep-ex]}
  \BibitemShut {NoStop}%
\bibitem [{\citenamefont {Bennett}\ \emph {et~al.}(2006)\citenamefont {Bennett}
  \emph {et~al.}}]{Bennett:2006fi}%
  \BibitemOpen
  \bibfield  {author} {\bibinfo {author} {\bibfnamefont {G.~W.}\ \bibnamefont
  {Bennett}} \emph {et~al.} (\bibinfo {collaboration} {Muon g-2}),\ }\href
  {\doibase 10.1103/PhysRevD.73.072003} {\bibfield  {journal} {\bibinfo
  {journal} {Phys. Rev.}\ }\textbf {\bibinfo {volume} {D73}},\ \bibinfo {pages}
  {072003} (\bibinfo {year} {2006})},\ \Eprint
  {http://arxiv.org/abs/hep-ex/0602035} {arXiv:hep-ex/0602035 [hep-ex]}
  \BibitemShut {NoStop}%
\bibitem [{\citenamefont {Crivellin}\ \emph {et~al.}(2020)\citenamefont
  {Crivellin}, \citenamefont {Hoferichter}, \citenamefont {Manzari},\ and\
  \citenamefont {Montull}}]{Crivellin:2020zul}%
  \BibitemOpen
  \bibfield  {author} {\bibinfo {author} {\bibfnamefont {A.}~\bibnamefont
  {Crivellin}}, \bibinfo {author} {\bibfnamefont {M.}~\bibnamefont
  {Hoferichter}}, \bibinfo {author} {\bibfnamefont {C.~A.}\ \bibnamefont
  {Manzari}}, \ and\ \bibinfo {author} {\bibfnamefont {M.}~\bibnamefont
  {Montull}},\ }\href@noop {} {\  (\bibinfo {year} {2020})},\ \Eprint
  {http://arxiv.org/abs/2003.04886} {arXiv:2003.04886 [hep-ph]} \BibitemShut
  {NoStop}%
\bibitem [{\citenamefont {Borsanyi}\ \emph {et~al.}(2020)\citenamefont
  {Borsanyi} \emph {et~al.}}]{Borsanyi:2020mff}%
  \BibitemOpen
  \bibfield  {author} {\bibinfo {author} {\bibfnamefont {S.}~\bibnamefont
  {Borsanyi}} \emph {et~al.},\ }\href@noop {} {\  (\bibinfo {year} {2020})},\
  \Eprint {http://arxiv.org/abs/2002.12347} {arXiv:2002.12347 [hep-lat]}
  \BibitemShut {NoStop}%
\bibitem [{\citenamefont {Prades}\ \emph {et~al.}(2009)\citenamefont {Prades},
  \citenamefont {de~Rafael},\ and\ \citenamefont {Vainshtein}}]{Prades:2009tw}%
  \BibitemOpen
  \bibfield  {author} {\bibinfo {author} {\bibfnamefont {J.}~\bibnamefont
  {Prades}}, \bibinfo {author} {\bibfnamefont {E.}~\bibnamefont {de~Rafael}}, \
  and\ \bibinfo {author} {\bibfnamefont {A.}~\bibnamefont {Vainshtein}},\
  }\href {\doibase 10.1142/9789814271844_0009} {\bibfield  {journal} {\bibinfo
  {journal} {Adv. Ser. Direct. High Energy Phys.}\ }\textbf {\bibinfo {volume}
  {20}},\ \bibinfo {pages} {303} (\bibinfo {year} {2009})},\ \Eprint
  {http://arxiv.org/abs/0901.0306} {arXiv:0901.0306 [hep-ph]} \BibitemShut
  {NoStop}%
\bibitem [{\citenamefont {Tanabashi}\ \emph {et~al.}(2018)\citenamefont
  {Tanabashi} \emph {et~al.}}]{Tanabashi:2018oca}%
  \BibitemOpen
  \bibfield  {author} {\bibinfo {author} {\bibfnamefont {M.}~\bibnamefont
  {Tanabashi}} \emph {et~al.} (\bibinfo {collaboration} {Particle Data
  Group}),\ }\href {\doibase 10.1103/PhysRevD.98.030001} {\bibfield  {journal}
  {\bibinfo  {journal} {Phys. Rev.}\ }\textbf {\bibinfo {volume} {D98}},\
  \bibinfo {pages} {030001} (\bibinfo {year} {2018})}\BibitemShut {NoStop}%
\bibitem [{\citenamefont {Benayoun}\ \emph {et~al.}(2013)\citenamefont
  {Benayoun}, \citenamefont {David}, \citenamefont {DelBuono},\ and\
  \citenamefont {Jegerlehner}}]{Benayoun:2012wc}%
  \BibitemOpen
  \bibfield  {author} {\bibinfo {author} {\bibfnamefont {M.}~\bibnamefont
  {Benayoun}}, \bibinfo {author} {\bibfnamefont {P.}~\bibnamefont {David}},
  \bibinfo {author} {\bibfnamefont {L.}~\bibnamefont {DelBuono}}, \ and\
  \bibinfo {author} {\bibfnamefont {F.}~\bibnamefont {Jegerlehner}},\ }\href
  {\doibase 10.1140/epjc/s10052-013-2453-3} {\bibfield  {journal} {\bibinfo
  {journal} {Eur. Phys. J.}\ }\textbf {\bibinfo {volume} {C73}},\ \bibinfo
  {pages} {2453} (\bibinfo {year} {2013})},\ \Eprint
  {http://arxiv.org/abs/1210.7184} {arXiv:1210.7184 [hep-ph]} \BibitemShut
  {NoStop}%
\bibitem [{\citenamefont {Blum}\ \emph {et~al.}(2013)\citenamefont {Blum},
  \citenamefont {Denig}, \citenamefont {Logashenko}, \citenamefont {de~Rafael},
  \citenamefont {Roberts}, \citenamefont {Teubner},\ and\ \citenamefont
  {Venanzoni}}]{Blum:2013xva}%
  \BibitemOpen
  \bibfield  {author} {\bibinfo {author} {\bibfnamefont {T.}~\bibnamefont
  {Blum}}, \bibinfo {author} {\bibfnamefont {A.}~\bibnamefont {Denig}},
  \bibinfo {author} {\bibfnamefont {I.}~\bibnamefont {Logashenko}}, \bibinfo
  {author} {\bibfnamefont {E.}~\bibnamefont {de~Rafael}}, \bibinfo {author}
  {\bibfnamefont {B.~L.}\ \bibnamefont {Roberts}}, \bibinfo {author}
  {\bibfnamefont {T.}~\bibnamefont {Teubner}}, \ and\ \bibinfo {author}
  {\bibfnamefont {G.}~\bibnamefont {Venanzoni}},\ }\href@noop {} {\  (\bibinfo
  {year} {2013})},\ \Eprint {http://arxiv.org/abs/1311.2198} {arXiv:1311.2198
  [hep-ph]} \BibitemShut {NoStop}%
\bibitem [{\citenamefont {Benayoun}\ \emph {et~al.}(2015)\citenamefont
  {Benayoun}, \citenamefont {David}, \citenamefont {DelBuono},\ and\
  \citenamefont {Jegerlehner}}]{Benayoun:2015gxa}%
  \BibitemOpen
  \bibfield  {author} {\bibinfo {author} {\bibfnamefont {M.}~\bibnamefont
  {Benayoun}}, \bibinfo {author} {\bibfnamefont {P.}~\bibnamefont {David}},
  \bibinfo {author} {\bibfnamefont {L.}~\bibnamefont {DelBuono}}, \ and\
  \bibinfo {author} {\bibfnamefont {F.}~\bibnamefont {Jegerlehner}},\ }\href
  {\doibase 10.1140/epjc/s10052-015-3830-x} {\bibfield  {journal} {\bibinfo
  {journal} {Eur. Phys. J.}\ }\textbf {\bibinfo {volume} {C75}},\ \bibinfo
  {pages} {613} (\bibinfo {year} {2015})},\ \Eprint
  {http://arxiv.org/abs/1507.02943} {arXiv:1507.02943 [hep-ph]} \BibitemShut
  {NoStop}%
\bibitem [{\citenamefont {Jegerlehner}(2018)}]{Jegerlehner:2017lbd}%
  \BibitemOpen
  \bibfield  {author} {\bibinfo {author} {\bibfnamefont {F.}~\bibnamefont
  {Jegerlehner}},\ }\bibfield  {booktitle} {\emph {\bibinfo {booktitle}
  {{Proceedings, KLOE-2 Workshop on $e^+ e^-$ Collision Physics at 1 GeV:
  Frascati, Italy, October $26-28$, 2016}}},\ }\href {\doibase
  10.1051/epjconf/201816600022} {\bibfield  {journal} {\bibinfo  {journal} {EPJ
  Web Conf.}\ }\textbf {\bibinfo {volume} {166}},\ \bibinfo {pages} {00022}
  (\bibinfo {year} {2018})},\ \Eprint {http://arxiv.org/abs/1705.00263}
  {arXiv:1705.00263 [hep-ph]} \BibitemShut {NoStop}%
\bibitem [{\citenamefont {Keshavarzi}\ \emph {et~al.}(2018)\citenamefont
  {Keshavarzi}, \citenamefont {Nomura},\ and\ \citenamefont
  {Teubner}}]{Keshavarzi:2018mgv}%
  \BibitemOpen
  \bibfield  {author} {\bibinfo {author} {\bibfnamefont {A.}~\bibnamefont
  {Keshavarzi}}, \bibinfo {author} {\bibfnamefont {D.}~\bibnamefont {Nomura}},
  \ and\ \bibinfo {author} {\bibfnamefont {T.}~\bibnamefont {Teubner}},\ }\href
  {\doibase 10.1103/PhysRevD.97.114025} {\bibfield  {journal} {\bibinfo
  {journal} {Phys. Rev.}\ }\textbf {\bibinfo {volume} {D97}},\ \bibinfo {pages}
  {114025} (\bibinfo {year} {2018})},\ \Eprint
  {http://arxiv.org/abs/1802.02995} {arXiv:1802.02995 [hep-ph]} \BibitemShut
  {NoStop}%
\bibitem [{\citenamefont {Grange}\ \emph {et~al.}(2015)\citenamefont {Grange}
  \emph {et~al.}}]{Grange:2015fou}%
  \BibitemOpen
  \bibfield  {author} {\bibinfo {author} {\bibfnamefont {J.}~\bibnamefont
  {Grange}} \emph {et~al.} (\bibinfo {collaboration} {Muon g-2}),\ }\href@noop
  {} {\  (\bibinfo {year} {2015})},\ \Eprint {http://arxiv.org/abs/1501.06858}
  {arXiv:1501.06858 [physics.ins-det]} \BibitemShut {NoStop}%
\bibitem [{\citenamefont {Al-Binni}\ \emph {et~al.}(2013)\citenamefont
  {Al-Binni} \emph {et~al.}}]{Kronfeld:2013uoa}%
  \BibitemOpen
  \bibfield  {author} {\bibinfo {author} {\bibfnamefont {U.}~\bibnamefont
  {Al-Binni}} \emph {et~al.},\ }in\ \href
  {http://www.slac.stanford.edu/econf/C1307292/docs/submittedArxivFiles/1306.5009.pdf}
  {\emph {\bibinfo {booktitle} {{Proceedings, 2013 Community Summer Study on
  the Future of U.S. Particle Physics: Snowmass on the Mississippi (CSS2013):
  Minneapolis, MN, USA, July 29-August 6, 2013}}}},\ \bibinfo {editor} {edited
  by\ \bibinfo {editor} {\bibfnamefont {A.~S.}\ \bibnamefont {Kronfeld}}\ and\
  \bibinfo {editor} {\bibfnamefont {R.~S.}\ \bibnamefont {Tschirhart}}}\
  (\bibinfo {year} {2013})\ \Eprint {http://arxiv.org/abs/1306.5009}
  {arXiv:1306.5009 [hep-ex]} \BibitemShut {NoStop}%
\bibitem [{\citenamefont {Abe}\ \emph {et~al.}(2019)\citenamefont {Abe} \emph
  {et~al.}}]{Abe:2019thb}%
  \BibitemOpen
  \bibfield  {author} {\bibinfo {author} {\bibfnamefont {M.}~\bibnamefont
  {Abe}} \emph {et~al.},\ }\href {\doibase 10.1093/ptep/ptz030} {\bibfield
  {journal} {\bibinfo  {journal} {PTEP}\ }\textbf {\bibinfo {volume} {2019}},\
  \bibinfo {pages} {053C02} (\bibinfo {year} {2019})},\ \Eprint
  {http://arxiv.org/abs/1901.03047} {arXiv:1901.03047 [physics.ins-det]}
  \BibitemShut {NoStop}%
\bibitem [{\citenamefont {Padley}\ \emph {et~al.}(2015)\citenamefont {Padley},
  \citenamefont {Sinha},\ and\ \citenamefont {Wang}}]{Padley:2015uma}%
  \BibitemOpen
  \bibfield  {author} {\bibinfo {author} {\bibfnamefont {B.~P.}\ \bibnamefont
  {Padley}}, \bibinfo {author} {\bibfnamefont {K.}~\bibnamefont {Sinha}}, \
  and\ \bibinfo {author} {\bibfnamefont {K.}~\bibnamefont {Wang}},\ }\href
  {\doibase 10.1103/PhysRevD.92.055025} {\bibfield  {journal} {\bibinfo
  {journal} {Phys. Rev.}\ }\textbf {\bibinfo {volume} {D92}},\ \bibinfo {pages}
  {055025} (\bibinfo {year} {2015})},\ \Eprint
  {http://arxiv.org/abs/1505.05877} {arXiv:1505.05877 [hep-ph]} \BibitemShut
  {NoStop}%
\bibitem [{\citenamefont {Yamaguchi}\ and\ \citenamefont
  {Yin}(2018)}]{Yamaguchi:2016oqz}%
  \BibitemOpen
  \bibfield  {author} {\bibinfo {author} {\bibfnamefont {M.}~\bibnamefont
  {Yamaguchi}}\ and\ \bibinfo {author} {\bibfnamefont {W.}~\bibnamefont
  {Yin}},\ }\href {\doibase 10.1093/ptep/pty002} {\bibfield  {journal}
  {\bibinfo  {journal} {PTEP}\ }\textbf {\bibinfo {volume} {2018}},\ \bibinfo
  {pages} {023B06} (\bibinfo {year} {2018})},\ \Eprint
  {http://arxiv.org/abs/1606.04953} {arXiv:1606.04953 [hep-ph]} \BibitemShut
  {NoStop}%
\bibitem [{\citenamefont {Yin}\ and\ \citenamefont
  {Yokozaki}(2016)}]{Yin:2016shg}%
  \BibitemOpen
  \bibfield  {author} {\bibinfo {author} {\bibfnamefont {W.}~\bibnamefont
  {Yin}}\ and\ \bibinfo {author} {\bibfnamefont {N.}~\bibnamefont {Yokozaki}},\
  }\href {\doibase 10.1016/j.physletb.2016.09.024} {\bibfield  {journal}
  {\bibinfo  {journal} {Phys. Lett.}\ }\textbf {\bibinfo {volume} {B762}},\
  \bibinfo {pages} {72} (\bibinfo {year} {2016})},\ \Eprint
  {http://arxiv.org/abs/1607.05705} {arXiv:1607.05705 [hep-ph]} \BibitemShut
  {NoStop}%
\bibitem [{\citenamefont {Endo}\ and\ \citenamefont
  {Yin}(2019)}]{Endo:2019bcj}%
  \BibitemOpen
  \bibfield  {author} {\bibinfo {author} {\bibfnamefont {M.}~\bibnamefont
  {Endo}}\ and\ \bibinfo {author} {\bibfnamefont {W.}~\bibnamefont {Yin}},\
  }\href {\doibase 10.1007/JHEP08(2019)122} {\bibfield  {journal} {\bibinfo
  {journal} {JHEP}\ }\textbf {\bibinfo {volume} {08}},\ \bibinfo {pages} {122}
  (\bibinfo {year} {2019})},\ \Eprint {http://arxiv.org/abs/1906.08768}
  {arXiv:1906.08768 [hep-ph]} \BibitemShut {NoStop}%
\bibitem [{\citenamefont {Endo}\ \emph {et~al.}(2020)\citenamefont {Endo},
  \citenamefont {Hamaguchi}, \citenamefont {Iwamoto},\ and\ \citenamefont
  {Kitahara}}]{Endo:2020mqz}%
  \BibitemOpen
  \bibfield  {author} {\bibinfo {author} {\bibfnamefont {M.}~\bibnamefont
  {Endo}}, \bibinfo {author} {\bibfnamefont {K.}~\bibnamefont {Hamaguchi}},
  \bibinfo {author} {\bibfnamefont {S.}~\bibnamefont {Iwamoto}}, \ and\
  \bibinfo {author} {\bibfnamefont {T.}~\bibnamefont {Kitahara}},\ }\href@noop
  {} {\  (\bibinfo {year} {2020})},\ \Eprint {http://arxiv.org/abs/2001.11025}
  {arXiv:2001.11025 [hep-ph]} \BibitemShut {NoStop}%
\bibitem [{\citenamefont {Endo}\ and\ \citenamefont
  {Ueda}(2019)}]{Endo:2019mxw}%
  \BibitemOpen
  \bibfield  {author} {\bibinfo {author} {\bibfnamefont {M.}~\bibnamefont
  {Endo}}\ and\ \bibinfo {author} {\bibfnamefont {D.}~\bibnamefont {Ueda}},\
  }\href@noop {} {\  (\bibinfo {year} {2019})},\ \Eprint
  {http://arxiv.org/abs/1911.10805} {arXiv:1911.10805 [hep-ph]} \BibitemShut
  {NoStop}%
\bibitem [{\citenamefont {Cox}\ \emph {et~al.}(2019)\citenamefont {Cox},
  \citenamefont {Han}, \citenamefont {Yanagida},\ and\ \citenamefont
  {Yokozaki}}]{Cox:2018vsv}%
  \BibitemOpen
  \bibfield  {author} {\bibinfo {author} {\bibfnamefont {P.}~\bibnamefont
  {Cox}}, \bibinfo {author} {\bibfnamefont {C.}~\bibnamefont {Han}}, \bibinfo
  {author} {\bibfnamefont {T.~T.}\ \bibnamefont {Yanagida}}, \ and\ \bibinfo
  {author} {\bibfnamefont {N.}~\bibnamefont {Yokozaki}},\ }\href {\doibase
  10.1007/JHEP08(2019)097} {\bibfield  {journal} {\bibinfo  {journal} {JHEP}\
  }\textbf {\bibinfo {volume} {08}},\ \bibinfo {pages} {097} (\bibinfo {year}
  {2019})},\ \Eprint {http://arxiv.org/abs/1811.12699} {arXiv:1811.12699
  [hep-ph]} \BibitemShut {NoStop}%
\bibitem [{\citenamefont {Konar}\ \emph {et~al.}(2018)\citenamefont {Konar},
  \citenamefont {Mondal},\ and\ \citenamefont {Swain}}]{Konar:2017oah}%
  \BibitemOpen
  \bibfield  {author} {\bibinfo {author} {\bibfnamefont {P.}~\bibnamefont
  {Konar}}, \bibinfo {author} {\bibfnamefont {T.}~\bibnamefont {Mondal}}, \
  and\ \bibinfo {author} {\bibfnamefont {A.~K.}\ \bibnamefont {Swain}},\ }\href
  {\doibase 10.1007/JHEP04(2018)024} {\bibfield  {journal} {\bibinfo  {journal}
  {JHEP}\ }\textbf {\bibinfo {volume} {04}},\ \bibinfo {pages} {024} (\bibinfo
  {year} {2018})},\ \Eprint {http://arxiv.org/abs/1710.08664} {arXiv:1710.08664
  [hep-ph]} \BibitemShut {NoStop}%
\bibitem [{\citenamefont {Un}\ and\ \citenamefont {Ozdal}(2016)}]{Un:2016hji}%
  \BibitemOpen
  \bibfield  {author} {\bibinfo {author} {\bibfnamefont {C.~S.}\ \bibnamefont
  {Un}}\ and\ \bibinfo {author} {\bibfnamefont {O.}~\bibnamefont {Ozdal}},\
  }\href {\doibase 10.1103/PhysRevD.93.055024} {\bibfield  {journal} {\bibinfo
  {journal} {Phys. Rev.}\ }\textbf {\bibinfo {volume} {D93}},\ \bibinfo {pages}
  {055024} (\bibinfo {year} {2016})},\ \Eprint
  {http://arxiv.org/abs/1601.02494} {arXiv:1601.02494 [hep-ph]} \BibitemShut
  {NoStop}%
\bibitem [{\citenamefont {An}\ and\ \citenamefont {Wang}(2015)}]{An:2015uwa}%
  \BibitemOpen
  \bibfield  {author} {\bibinfo {author} {\bibfnamefont {H.}~\bibnamefont
  {An}}\ and\ \bibinfo {author} {\bibfnamefont {L.-T.}\ \bibnamefont {Wang}},\
  }\href {\doibase 10.1103/PhysRevLett.115.181602} {\bibfield  {journal}
  {\bibinfo  {journal} {Phys. Rev. Lett.}\ }\textbf {\bibinfo {volume} {115}},\
  \bibinfo {pages} {181602} (\bibinfo {year} {2015})},\ \Eprint
  {http://arxiv.org/abs/1506.00653} {arXiv:1506.00653 [hep-ph]} \BibitemShut
  {NoStop}%
\bibitem [{\citenamefont {Chakraborty}\ and\ \citenamefont
  {Chakraborty}(2016)}]{Chakraborty:2015bsk}%
  \BibitemOpen
  \bibfield  {author} {\bibinfo {author} {\bibfnamefont {A.}~\bibnamefont
  {Chakraborty}}\ and\ \bibinfo {author} {\bibfnamefont {S.}~\bibnamefont
  {Chakraborty}},\ }\href {\doibase 10.1103/PhysRevD.93.075035} {\bibfield
  {journal} {\bibinfo  {journal} {Phys. Rev.}\ }\textbf {\bibinfo {volume}
  {D93}},\ \bibinfo {pages} {075035} (\bibinfo {year} {2016})},\ \Eprint
  {http://arxiv.org/abs/1511.08874} {arXiv:1511.08874 [hep-ph]} \BibitemShut
  {NoStop}%
\bibitem [{\citenamefont {Jana}\ \emph {et~al.}(2020)\citenamefont {Jana},
  \citenamefont {K.},\ and\ \citenamefont {Saad}}]{Jana:2020pxx}%
  \BibitemOpen
  \bibfield  {author} {\bibinfo {author} {\bibfnamefont {S.}~\bibnamefont
  {Jana}}, \bibinfo {author} {\bibfnamefont {V.~P.}\ \bibnamefont {K.}}, \ and\
  \bibinfo {author} {\bibfnamefont {S.}~\bibnamefont {Saad}},\ }\href@noop {}
  {\  (\bibinfo {year} {2020})},\ \Eprint {http://arxiv.org/abs/2003.03386}
  {arXiv:2003.03386 [hep-ph]} \BibitemShut {NoStop}%
\bibitem [{\citenamefont {Lindner}\ \emph {et~al.}(2018)\citenamefont
  {Lindner}, \citenamefont {Platscher},\ and\ \citenamefont
  {Queiroz}}]{Lindner:2016bgg}%
  \BibitemOpen
  \bibfield  {author} {\bibinfo {author} {\bibfnamefont {M.}~\bibnamefont
  {Lindner}}, \bibinfo {author} {\bibfnamefont {M.}~\bibnamefont {Platscher}},
  \ and\ \bibinfo {author} {\bibfnamefont {F.~S.}\ \bibnamefont {Queiroz}},\
  }\href {\doibase 10.1016/j.physrep.2017.12.001} {\bibfield  {journal}
  {\bibinfo  {journal} {Phys. Rept.}\ }\textbf {\bibinfo {volume} {731}},\
  \bibinfo {pages} {1} (\bibinfo {year} {2018})},\ \Eprint
  {http://arxiv.org/abs/1610.06587} {arXiv:1610.06587 [hep-ph]} \BibitemShut
  {NoStop}%
\bibitem [{\citenamefont {Dermisek}\ and\ \citenamefont
  {Raval}(2013)}]{Dermisek:2013gta}%
  \BibitemOpen
  \bibfield  {author} {\bibinfo {author} {\bibfnamefont {R.}~\bibnamefont
  {Dermisek}}\ and\ \bibinfo {author} {\bibfnamefont {A.}~\bibnamefont
  {Raval}},\ }\href {\doibase 10.1103/PhysRevD.88.013017} {\bibfield  {journal}
  {\bibinfo  {journal} {Phys. Rev.}\ }\textbf {\bibinfo {volume} {D88}},\
  \bibinfo {pages} {013017} (\bibinfo {year} {2013})},\ \Eprint
  {http://arxiv.org/abs/1305.3522} {arXiv:1305.3522 [hep-ph]} \BibitemShut
  {NoStop}%
\bibitem [{\citenamefont {Poh}\ and\ \citenamefont {Raby}(2017)}]{Poh:2017tfo}%
  \BibitemOpen
  \bibfield  {author} {\bibinfo {author} {\bibfnamefont {Z.}~\bibnamefont
  {Poh}}\ and\ \bibinfo {author} {\bibfnamefont {S.}~\bibnamefont {Raby}},\
  }\href {\doibase 10.1103/PhysRevD.96.015032} {\bibfield  {journal} {\bibinfo
  {journal} {Phys. Rev.}\ }\textbf {\bibinfo {volume} {D96}},\ \bibinfo {pages}
  {015032} (\bibinfo {year} {2017})},\ \Eprint
  {http://arxiv.org/abs/1705.07007} {arXiv:1705.07007 [hep-ph]} \BibitemShut
  {NoStop}%
\bibitem [{\citenamefont {Barman}\ \emph {et~al.}(2019)\citenamefont {Barman},
  \citenamefont {Borah}, \citenamefont {Mukherjee},\ and\ \citenamefont
  {Nandi}}]{Barman:2018jhz}%
  \BibitemOpen
  \bibfield  {author} {\bibinfo {author} {\bibfnamefont {B.}~\bibnamefont
  {Barman}}, \bibinfo {author} {\bibfnamefont {D.}~\bibnamefont {Borah}},
  \bibinfo {author} {\bibfnamefont {L.}~\bibnamefont {Mukherjee}}, \ and\
  \bibinfo {author} {\bibfnamefont {S.}~\bibnamefont {Nandi}},\ }\href
  {\doibase 10.1103/PhysRevD.100.115010} {\bibfield  {journal} {\bibinfo
  {journal} {Phys. Rev.}\ }\textbf {\bibinfo {volume} {D100}},\ \bibinfo
  {pages} {115010} (\bibinfo {year} {2019})},\ \Eprint
  {http://arxiv.org/abs/1808.06639} {arXiv:1808.06639 [hep-ph]} \BibitemShut
  {NoStop}%
\bibitem [{\citenamefont {Pisano}\ and\ \citenamefont
  {Pleitez}(1992)}]{Pisano:1991ee}%
  \BibitemOpen
  \bibfield  {author} {\bibinfo {author} {\bibfnamefont {F.}~\bibnamefont
  {Pisano}}\ and\ \bibinfo {author} {\bibfnamefont {V.}~\bibnamefont
  {Pleitez}},\ }\href {\doibase 10.1103/PhysRevD.46.410} {\bibfield  {journal}
  {\bibinfo  {journal} {Phys. Rev.}\ }\textbf {\bibinfo {volume} {D46}},\
  \bibinfo {pages} {410} (\bibinfo {year} {1992})},\ \Eprint
  {http://arxiv.org/abs/hep-ph/9206242} {arXiv:hep-ph/9206242 [hep-ph]}
  \BibitemShut {NoStop}%
\bibitem [{\citenamefont {Foot}\ \emph {et~al.}(1993)\citenamefont {Foot},
  \citenamefont {Hernandez}, \citenamefont {Pisano},\ and\ \citenamefont
  {Pleitez}}]{Foot:1992rh}%
  \BibitemOpen
  \bibfield  {author} {\bibinfo {author} {\bibfnamefont {R.}~\bibnamefont
  {Foot}}, \bibinfo {author} {\bibfnamefont {O.~F.}\ \bibnamefont {Hernandez}},
  \bibinfo {author} {\bibfnamefont {F.}~\bibnamefont {Pisano}}, \ and\ \bibinfo
  {author} {\bibfnamefont {V.}~\bibnamefont {Pleitez}},\ }\href {\doibase
  10.1103/PhysRevD.47.4158} {\bibfield  {journal} {\bibinfo  {journal} {Phys.
  Rev.}\ }\textbf {\bibinfo {volume} {D47}},\ \bibinfo {pages} {4158} (\bibinfo
  {year} {1993})},\ \Eprint {http://arxiv.org/abs/hep-ph/9207264}
  {arXiv:hep-ph/9207264 [hep-ph]} \BibitemShut {NoStop}%
\bibitem [{\citenamefont {Long}(1996{\natexlab{a}})}]{Hoang:1995vq}%
  \BibitemOpen
  \bibfield  {author} {\bibinfo {author} {\bibfnamefont {H.~N.}\ \bibnamefont
  {Long}},\ }\href {\doibase 10.1103/PhysRevD.53.437} {\bibfield  {journal}
  {\bibinfo  {journal} {Phys. Rev.}\ }\textbf {\bibinfo {volume} {D53}},\
  \bibinfo {pages} {437} (\bibinfo {year} {1996}{\natexlab{a}})},\ \Eprint
  {http://arxiv.org/abs/hep-ph/9504274} {arXiv:hep-ph/9504274 [hep-ph]}
  \BibitemShut {NoStop}%
\bibitem [{\citenamefont {Baldini}\ \emph {et~al.}(2016)\citenamefont {Baldini}
  \emph {et~al.}}]{TheMEG:2016wtm}%
  \BibitemOpen
  \bibfield  {author} {\bibinfo {author} {\bibfnamefont {A.~M.}\ \bibnamefont
  {Baldini}} \emph {et~al.} (\bibinfo {collaboration} {MEG}),\ }\href {\doibase
  10.1140/epjc/s10052-016-4271-x} {\bibfield  {journal} {\bibinfo  {journal}
  {Eur. Phys. J.}\ }\textbf {\bibinfo {volume} {C76}},\ \bibinfo {pages} {434}
  (\bibinfo {year} {2016})},\ \Eprint {http://arxiv.org/abs/1605.05081}
  {arXiv:1605.05081 [hep-ex]} \BibitemShut {NoStop}%
\bibitem [{\citenamefont {Mori}(2017)}]{Mori:2016vwi}%
  \BibitemOpen
  \bibfield  {author} {\bibinfo {author} {\bibfnamefont {T.}~\bibnamefont
  {Mori}} (\bibinfo {collaboration} {MEG}),\ }\bibfield  {booktitle} {\emph
  {\bibinfo {booktitle} {{Proceedings, 30th Rencontres de Physique de La
  Vallée d'Aoste: La Thuile, Aosta Valley, Italy, March 6-12, 2016}}},\ }\href
  {\doibase 10.1393/ncc/i2016-16325-7} {\bibfield  {journal} {\bibinfo
  {journal} {Nuovo Cim.}\ }\textbf {\bibinfo {volume} {C39}},\ \bibinfo {pages}
  {325} (\bibinfo {year} {2017})},\ \Eprint {http://arxiv.org/abs/1606.08168}
  {arXiv:1606.08168 [hep-ex]} \BibitemShut {NoStop}%
\bibitem [{\citenamefont {Freitas}\ and\ \citenamefont
  {Westhoff}(2014)}]{Freitas:2014jla}%
  \BibitemOpen
  \bibfield  {author} {\bibinfo {author} {\bibfnamefont {A.}~\bibnamefont
  {Freitas}}\ and\ \bibinfo {author} {\bibfnamefont {S.}~\bibnamefont
  {Westhoff}},\ }\href {\doibase 10.1007/JHEP10(2014)116} {\bibfield  {journal}
  {\bibinfo  {journal} {JHEP}\ }\textbf {\bibinfo {volume} {10}},\ \bibinfo
  {pages} {116} (\bibinfo {year} {2014})},\ \Eprint
  {http://arxiv.org/abs/1408.1959} {arXiv:1408.1959 [hep-ph]} \BibitemShut
  {NoStop}%
\bibitem [{\citenamefont {Wang}\ and\ \citenamefont
  {Han}(2015)}]{Wang:2014sda}%
  \BibitemOpen
  \bibfield  {author} {\bibinfo {author} {\bibfnamefont {L.}~\bibnamefont
  {Wang}}\ and\ \bibinfo {author} {\bibfnamefont {X.-F.}\ \bibnamefont {Han}},\
  }\href {\doibase 10.1007/JHEP05(2015)039} {\bibfield  {journal} {\bibinfo
  {journal} {JHEP}\ }\textbf {\bibinfo {volume} {05}},\ \bibinfo {pages} {039}
  (\bibinfo {year} {2015})},\ \Eprint {http://arxiv.org/abs/1412.4874}
  {arXiv:1412.4874 [hep-ph]} \BibitemShut {NoStop}%
\bibitem [{\citenamefont {Hektor}\ \emph {et~al.}(2015)\citenamefont {Hektor},
  \citenamefont {Kannike},\ and\ \citenamefont {Marzola}}]{Hektor:2015zba}%
  \BibitemOpen
  \bibfield  {author} {\bibinfo {author} {\bibfnamefont {A.}~\bibnamefont
  {Hektor}}, \bibinfo {author} {\bibfnamefont {K.}~\bibnamefont {Kannike}}, \
  and\ \bibinfo {author} {\bibfnamefont {L.}~\bibnamefont {Marzola}},\ }\href
  {\doibase 10.1088/1475-7516/2015/10/025} {\bibfield  {journal} {\bibinfo
  {journal} {JCAP}\ }\textbf {\bibinfo {volume} {1510}},\ \bibinfo {pages}
  {025} (\bibinfo {year} {2015})},\ \Eprint {http://arxiv.org/abs/1507.05096}
  {arXiv:1507.05096 [hep-ph]} \BibitemShut {NoStop}%
\bibitem [{\citenamefont {Cherchiglia}\ \emph {et~al.}(2018)\citenamefont
  {Cherchiglia}, \citenamefont {Stöckinger},\ and\ \citenamefont
  {Stöckinger-Kim}}]{Cherchiglia:2017uwv}%
  \BibitemOpen
  \bibfield  {author} {\bibinfo {author} {\bibfnamefont {A.}~\bibnamefont
  {Cherchiglia}}, \bibinfo {author} {\bibfnamefont {D.}~\bibnamefont
  {Stöckinger}}, \ and\ \bibinfo {author} {\bibfnamefont {H.}~\bibnamefont
  {Stöckinger-Kim}},\ }\href {\doibase 10.1103/PhysRevD.98.035001} {\bibfield
  {journal} {\bibinfo  {journal} {Phys. Rev.}\ }\textbf {\bibinfo {volume}
  {D98}},\ \bibinfo {pages} {035001} (\bibinfo {year} {2018})},\ \Eprint
  {http://arxiv.org/abs/1711.11567} {arXiv:1711.11567 [hep-ph]} \BibitemShut
  {NoStop}%
\bibitem [{\citenamefont {Wang}\ \emph {et~al.}(2019)\citenamefont {Wang},
  \citenamefont {Yang}, \citenamefont {Zhang},\ and\ \citenamefont
  {Zhang}}]{Wang:2018hnw}%
  \BibitemOpen
  \bibfield  {author} {\bibinfo {author} {\bibfnamefont {L.}~\bibnamefont
  {Wang}}, \bibinfo {author} {\bibfnamefont {J.~M.}\ \bibnamefont {Yang}},
  \bibinfo {author} {\bibfnamefont {M.}~\bibnamefont {Zhang}}, \ and\ \bibinfo
  {author} {\bibfnamefont {Y.}~\bibnamefont {Zhang}},\ }\href {\doibase
  10.1016/j.physletb.2018.11.045} {\bibfield  {journal} {\bibinfo  {journal}
  {Phys. Lett.}\ }\textbf {\bibinfo {volume} {B788}},\ \bibinfo {pages} {519}
  (\bibinfo {year} {2019})},\ \Eprint {http://arxiv.org/abs/1809.05857}
  {arXiv:1809.05857 [hep-ph]} \BibitemShut {NoStop}%
\bibitem [{\citenamefont {Iguro}\ \emph {et~al.}(2019)\citenamefont {Iguro},
  \citenamefont {Omura},\ and\ \citenamefont {Takeuchi}}]{Iguro:2019sly}%
  \BibitemOpen
  \bibfield  {author} {\bibinfo {author} {\bibfnamefont {S.}~\bibnamefont
  {Iguro}}, \bibinfo {author} {\bibfnamefont {Y.}~\bibnamefont {Omura}}, \ and\
  \bibinfo {author} {\bibfnamefont {M.}~\bibnamefont {Takeuchi}},\ }\href
  {\doibase 10.1007/JHEP11(2019)130} {\bibfield  {journal} {\bibinfo  {journal}
  {JHEP}\ }\textbf {\bibinfo {volume} {11}},\ \bibinfo {pages} {130} (\bibinfo
  {year} {2019})},\ \Eprint {http://arxiv.org/abs/1907.09845} {arXiv:1907.09845
  [hep-ph]} \BibitemShut {NoStop}%
\bibitem [{\citenamefont {Queiroz}\ and\ \citenamefont
  {Shepherd}(2014)}]{Queiroz:2014zfa}%
  \BibitemOpen
  \bibfield  {author} {\bibinfo {author} {\bibfnamefont {F.~S.}\ \bibnamefont
  {Queiroz}}\ and\ \bibinfo {author} {\bibfnamefont {W.}~\bibnamefont
  {Shepherd}},\ }\href {\doibase 10.1103/PhysRevD.89.095024} {\bibfield
  {journal} {\bibinfo  {journal} {Phys. Rev.}\ }\textbf {\bibinfo {volume}
  {D89}},\ \bibinfo {pages} {095024} (\bibinfo {year} {2014})},\ \Eprint
  {http://arxiv.org/abs/1403.2309} {arXiv:1403.2309 [hep-ph]} \BibitemShut
  {NoStop}%
\bibitem [{\citenamefont {Altmannshofer}\ \emph {et~al.}(2014)\citenamefont
  {Altmannshofer}, \citenamefont {Bauer},\ and\ \citenamefont
  {Carena}}]{Altmannshofer:2013zba}%
  \BibitemOpen
  \bibfield  {author} {\bibinfo {author} {\bibfnamefont {W.}~\bibnamefont
  {Altmannshofer}}, \bibinfo {author} {\bibfnamefont {M.}~\bibnamefont
  {Bauer}}, \ and\ \bibinfo {author} {\bibfnamefont {M.}~\bibnamefont
  {Carena}},\ }\href {\doibase 10.1007/JHEP01(2014)060} {\bibfield  {journal}
  {\bibinfo  {journal} {JHEP}\ }\textbf {\bibinfo {volume} {01}},\ \bibinfo
  {pages} {060} (\bibinfo {year} {2014})},\ \Eprint
  {http://arxiv.org/abs/1308.1987} {arXiv:1308.1987 [hep-ph]} \BibitemShut
  {NoStop}%
\bibitem [{\citenamefont {Fregolente}\ and\ \citenamefont
  {Tonasse}(2003)}]{Fregolente:2002nx}%
  \BibitemOpen
  \bibfield  {author} {\bibinfo {author} {\bibfnamefont {D.}~\bibnamefont
  {Fregolente}}\ and\ \bibinfo {author} {\bibfnamefont {M.~D.}\ \bibnamefont
  {Tonasse}},\ }\href {\doibase 10.1016/S0370-2693(03)00037-6} {\bibfield
  {journal} {\bibinfo  {journal} {Phys. Lett.}\ }\textbf {\bibinfo {volume}
  {B555}},\ \bibinfo {pages} {7} (\bibinfo {year} {2003})},\ \Eprint
  {http://arxiv.org/abs/hep-ph/0209119} {arXiv:hep-ph/0209119 [hep-ph]}
  \BibitemShut {NoStop}%
\bibitem [{\citenamefont {Long}\ and\ \citenamefont
  {Lan}(2003)}]{Hoang:2003vj}%
  \BibitemOpen
  \bibfield  {author} {\bibinfo {author} {\bibfnamefont {H.~N.}\ \bibnamefont
  {Long}}\ and\ \bibinfo {author} {\bibfnamefont {N.~Q.}\ \bibnamefont {Lan}},\
  }\href {\doibase 10.1209/epl/i2003-00267-5} {\bibfield  {journal} {\bibinfo
  {journal} {Europhys. Lett.}\ }\textbf {\bibinfo {volume} {64}},\ \bibinfo
  {pages} {571} (\bibinfo {year} {2003})},\ \Eprint
  {http://arxiv.org/abs/hep-ph/0309038} {arXiv:hep-ph/0309038 [hep-ph]}
  \BibitemShut {NoStop}%
\bibitem [{\citenamefont {de~S.~Pires}\ and\ \citenamefont {Rodrigues~da
  Silva}(2007)}]{deS.Pires:2007gi}%
  \BibitemOpen
  \bibfield  {author} {\bibinfo {author} {\bibfnamefont {C.~A.}\ \bibnamefont
  {de~S.~Pires}}\ and\ \bibinfo {author} {\bibfnamefont {P.~S.}\ \bibnamefont
  {Rodrigues~da Silva}},\ }\href {\doibase 10.1088/1475-7516/2007/12/012}
  {\bibfield  {journal} {\bibinfo  {journal} {JCAP}\ }\textbf {\bibinfo
  {volume} {0712}},\ \bibinfo {pages} {012} (\bibinfo {year} {2007})},\ \Eprint
  {http://arxiv.org/abs/0710.2104} {arXiv:0710.2104 [hep-ph]} \BibitemShut
  {NoStop}%
\bibitem [{\citenamefont {Mizukoshi}\ \emph {et~al.}(2011)\citenamefont
  {Mizukoshi}, \citenamefont {de~S.~Pires}, \citenamefont {Queiroz},\ and\
  \citenamefont {Rodrigues~da Silva}}]{Mizukoshi:2010ky}%
  \BibitemOpen
  \bibfield  {author} {\bibinfo {author} {\bibfnamefont {J.~K.}\ \bibnamefont
  {Mizukoshi}}, \bibinfo {author} {\bibfnamefont {C.~A.}\ \bibnamefont
  {de~S.~Pires}}, \bibinfo {author} {\bibfnamefont {F.~S.}\ \bibnamefont
  {Queiroz}}, \ and\ \bibinfo {author} {\bibfnamefont {P.~S.}\ \bibnamefont
  {Rodrigues~da Silva}},\ }\href {\doibase 10.1103/PhysRevD.83.065024}
  {\bibfield  {journal} {\bibinfo  {journal} {Phys. Rev.}\ }\textbf {\bibinfo
  {volume} {D83}},\ \bibinfo {pages} {065024} (\bibinfo {year} {2011})},\
  \Eprint {http://arxiv.org/abs/1010.4097} {arXiv:1010.4097 [hep-ph]}
  \BibitemShut {NoStop}%
\bibitem [{\citenamefont {Profumo}\ and\ \citenamefont
  {Queiroz}(2014)}]{Profumo:2013sca}%
  \BibitemOpen
  \bibfield  {author} {\bibinfo {author} {\bibfnamefont {S.}~\bibnamefont
  {Profumo}}\ and\ \bibinfo {author} {\bibfnamefont {F.~S.}\ \bibnamefont
  {Queiroz}},\ }\href {\doibase 10.1140/epjc/s10052-014-2960-x} {\bibfield
  {journal} {\bibinfo  {journal} {Eur. Phys. J.}\ }\textbf {\bibinfo {volume}
  {C74}},\ \bibinfo {pages} {2960} (\bibinfo {year} {2014})},\ \Eprint
  {http://arxiv.org/abs/1307.7802} {arXiv:1307.7802 [hep-ph]} \BibitemShut
  {NoStop}%
\bibitem [{\citenamefont {Dong}\ \emph
  {et~al.}(2013{\natexlab{a}})\citenamefont {Dong}, \citenamefont {Nguyen},\
  and\ \citenamefont {Soa}}]{Dong:2013ioa}%
  \BibitemOpen
  \bibfield  {author} {\bibinfo {author} {\bibfnamefont {P.~V.}\ \bibnamefont
  {Dong}}, \bibinfo {author} {\bibfnamefont {T.~P.}\ \bibnamefont {Nguyen}}, \
  and\ \bibinfo {author} {\bibfnamefont {D.~V.}\ \bibnamefont {Soa}},\ }\href
  {\doibase 10.1103/PhysRevD.88.095014} {\bibfield  {journal} {\bibinfo
  {journal} {Phys. Rev.}\ }\textbf {\bibinfo {volume} {D88}},\ \bibinfo {pages}
  {095014} (\bibinfo {year} {2013}{\natexlab{a}})},\ \Eprint
  {http://arxiv.org/abs/1308.4097} {arXiv:1308.4097 [hep-ph]} \BibitemShut
  {NoStop}%
\bibitem [{\citenamefont {Dong}\ \emph
  {et~al.}(2013{\natexlab{b}})\citenamefont {Dong}, \citenamefont {Hung},\ and\
  \citenamefont {Tham}}]{Dong:2013wca}%
  \BibitemOpen
  \bibfield  {author} {\bibinfo {author} {\bibfnamefont {P.~V.}\ \bibnamefont
  {Dong}}, \bibinfo {author} {\bibfnamefont {H.~T.}\ \bibnamefont {Hung}}, \
  and\ \bibinfo {author} {\bibfnamefont {T.~D.}\ \bibnamefont {Tham}},\ }\href
  {\doibase 10.1103/PhysRevD.87.115003} {\bibfield  {journal} {\bibinfo
  {journal} {Phys. Rev.}\ }\textbf {\bibinfo {volume} {D87}},\ \bibinfo {pages}
  {115003} (\bibinfo {year} {2013}{\natexlab{b}})},\ \Eprint
  {http://arxiv.org/abs/1305.0369} {arXiv:1305.0369 [hep-ph]} \BibitemShut
  {NoStop}%
\bibitem [{\citenamefont {Cogollo}\ \emph {et~al.}(2014)\citenamefont
  {Cogollo}, \citenamefont {Gonzalez-Morales}, \citenamefont {Queiroz},\ and\
  \citenamefont {Teles}}]{Cogollo:2014jia}%
  \BibitemOpen
  \bibfield  {author} {\bibinfo {author} {\bibfnamefont {D.}~\bibnamefont
  {Cogollo}}, \bibinfo {author} {\bibfnamefont {A.~X.}\ \bibnamefont
  {Gonzalez-Morales}}, \bibinfo {author} {\bibfnamefont {F.~S.}\ \bibnamefont
  {Queiroz}}, \ and\ \bibinfo {author} {\bibfnamefont {P.~R.}\ \bibnamefont
  {Teles}},\ }\href {\doibase 10.1088/1475-7516/2014/11/002} {\bibfield
  {journal} {\bibinfo  {journal} {JCAP}\ }\textbf {\bibinfo {volume} {1411}},\
  \bibinfo {pages} {002} (\bibinfo {year} {2014})},\ \Eprint
  {http://arxiv.org/abs/1402.3271} {arXiv:1402.3271 [hep-ph]} \BibitemShut
  {NoStop}%
\bibitem [{\citenamefont {Dong}\ \emph
  {et~al.}(2014{\natexlab{a}})\citenamefont {Dong}, \citenamefont {Huong},
  \citenamefont {Queiroz},\ and\ \citenamefont {Thuy}}]{Dong:2014wsa}%
  \BibitemOpen
  \bibfield  {author} {\bibinfo {author} {\bibfnamefont {P.~V.}\ \bibnamefont
  {Dong}}, \bibinfo {author} {\bibfnamefont {D.~T.}\ \bibnamefont {Huong}},
  \bibinfo {author} {\bibfnamefont {F.~S.}\ \bibnamefont {Queiroz}}, \ and\
  \bibinfo {author} {\bibfnamefont {N.~T.}\ \bibnamefont {Thuy}},\ }\href
  {\doibase 10.1103/PhysRevD.90.075021} {\bibfield  {journal} {\bibinfo
  {journal} {Phys. Rev.}\ }\textbf {\bibinfo {volume} {D90}},\ \bibinfo {pages}
  {075021} (\bibinfo {year} {2014}{\natexlab{a}})},\ \Eprint
  {http://arxiv.org/abs/1405.2591} {arXiv:1405.2591 [hep-ph]} \BibitemShut
  {NoStop}%
\bibitem [{\citenamefont {Dong}\ \emph
  {et~al.}(2014{\natexlab{b}})\citenamefont {Dong}, \citenamefont {Ngan},\ and\
  \citenamefont {Soa}}]{Dong:2014esa}%
  \BibitemOpen
  \bibfield  {author} {\bibinfo {author} {\bibfnamefont {P.~V.}\ \bibnamefont
  {Dong}}, \bibinfo {author} {\bibfnamefont {N.~T.~K.}\ \bibnamefont {Ngan}}, \
  and\ \bibinfo {author} {\bibfnamefont {D.~V.}\ \bibnamefont {Soa}},\ }\href
  {\doibase 10.1103/PhysRevD.90.075019} {\bibfield  {journal} {\bibinfo
  {journal} {Phys. Rev.}\ }\textbf {\bibinfo {volume} {D90}},\ \bibinfo {pages}
  {075019} (\bibinfo {year} {2014}{\natexlab{b}})},\ \Eprint
  {http://arxiv.org/abs/1407.3839} {arXiv:1407.3839 [hep-ph]} \BibitemShut
  {NoStop}%
\bibitem [{\citenamefont {Carvajal}\ \emph {et~al.}(2017)\citenamefont
  {Carvajal}, \citenamefont {Sanchez-Vega},\ and\ \citenamefont
  {Zapata}}]{Carvajal:2017gjj}%
  \BibitemOpen
  \bibfield  {author} {\bibinfo {author} {\bibfnamefont {C.~D.~R.}\
  \bibnamefont {Carvajal}}, \bibinfo {author} {\bibfnamefont {B.~L.}\
  \bibnamefont {Sanchez-Vega}}, \ and\ \bibinfo {author} {\bibfnamefont
  {O.}~\bibnamefont {Zapata}},\ }\href {\doibase 10.1103/PhysRevD.96.115035}
  {\bibfield  {journal} {\bibinfo  {journal} {Phys. Rev.}\ }\textbf {\bibinfo
  {volume} {D96}},\ \bibinfo {pages} {115035} (\bibinfo {year} {2017})},\
  \Eprint {http://arxiv.org/abs/1704.08340} {arXiv:1704.08340 [hep-ph]}
  \BibitemShut {NoStop}%
\bibitem [{\citenamefont {Montero}\ \emph {et~al.}(2018)\citenamefont
  {Montero}, \citenamefont {Romero},\ and\ \citenamefont
  {Sanchez-Vega}}]{Montero:2017yvy}%
  \BibitemOpen
  \bibfield  {author} {\bibinfo {author} {\bibfnamefont {J.~C.}\ \bibnamefont
  {Montero}}, \bibinfo {author} {\bibfnamefont {A.}~\bibnamefont {Romero}}, \
  and\ \bibinfo {author} {\bibfnamefont {B.~L.}\ \bibnamefont {Sanchez-Vega}},\
  }\href {\doibase 10.1103/PhysRevD.97.063015} {\bibfield  {journal} {\bibinfo
  {journal} {Phys. Rev.}\ }\textbf {\bibinfo {volume} {D97}},\ \bibinfo {pages}
  {063015} (\bibinfo {year} {2018})},\ \Eprint
  {http://arxiv.org/abs/1709.04535} {arXiv:1709.04535 [hep-ph]} \BibitemShut
  {NoStop}%
\bibitem [{\citenamefont {Huong}\ \emph {et~al.}(2019)\citenamefont {Huong},
  \citenamefont {Dinh}, \citenamefont {Thien},\ and\ \citenamefont
  {Van~Dong}}]{Huong:2019vej}%
  \BibitemOpen
  \bibfield  {author} {\bibinfo {author} {\bibfnamefont {D.~T.}\ \bibnamefont
  {Huong}}, \bibinfo {author} {\bibfnamefont {D.~N.}\ \bibnamefont {Dinh}},
  \bibinfo {author} {\bibfnamefont {L.~D.}\ \bibnamefont {Thien}}, \ and\
  \bibinfo {author} {\bibfnamefont {P.}~\bibnamefont {Van~Dong}},\ }\href
  {\doibase 10.1007/JHEP08(2019)051} {\bibfield  {journal} {\bibinfo  {journal}
  {JHEP}\ }\textbf {\bibinfo {volume} {08}},\ \bibinfo {pages} {051} (\bibinfo
  {year} {2019})},\ \Eprint {http://arxiv.org/abs/1906.05240} {arXiv:1906.05240
  [hep-ph]} \BibitemShut {NoStop}%
\bibitem [{\citenamefont {Cabarcas}\ \emph {et~al.}(2012)\citenamefont
  {Cabarcas}, \citenamefont {Duarte},\ and\ \citenamefont
  {Rodriguez}}]{Cabarcas:2012uf}%
  \BibitemOpen
  \bibfield  {author} {\bibinfo {author} {\bibfnamefont {J.~M.}\ \bibnamefont
  {Cabarcas}}, \bibinfo {author} {\bibfnamefont {J.}~\bibnamefont {Duarte}}, \
  and\ \bibinfo {author} {\bibfnamefont {J.~A.}\ \bibnamefont {Rodriguez}},\
  }\bibfield  {booktitle} {\emph {\bibinfo {booktitle} {{Proceedings on 11th
  International Conference on Heavy Quarks and Leptons (HQL 2012): Prague,
  Czech Republic, Jun. 11-15, 2012}}},\ }\href {\doibase 10.22323/1.166.0072}
  {\bibfield  {journal} {\bibinfo  {journal} {PoS}\ }\textbf {\bibinfo {volume}
  {HQL2012}},\ \bibinfo {pages} {072} (\bibinfo {year} {2012})},\ \Eprint
  {http://arxiv.org/abs/1212.3586} {arXiv:1212.3586 [hep-ph]} \BibitemShut
  {NoStop}%
\bibitem [{\citenamefont {Santos}\ and\ \citenamefont
  {Vasconcelos}(2018)}]{Santos:2017jbv}%
  \BibitemOpen
  \bibfield  {author} {\bibinfo {author} {\bibfnamefont {A.~C.~O.}\
  \bibnamefont {Santos}}\ and\ \bibinfo {author} {\bibfnamefont
  {P.}~\bibnamefont {Vasconcelos}},\ }\href {\doibase 10.1155/2018/9132381}
  {\bibfield  {journal} {\bibinfo  {journal} {Adv. High Energy Phys.}\ }\textbf
  {\bibinfo {volume} {2018}},\ \bibinfo {pages} {9132381} (\bibinfo {year}
  {2018})},\ \Eprint {http://arxiv.org/abs/1708.03955} {arXiv:1708.03955
  [hep-ph]} \BibitemShut {NoStop}%
\bibitem [{\citenamefont {Barreto}\ \emph {et~al.}(2018)\citenamefont
  {Barreto}, \citenamefont {Dias}, \citenamefont {Leite}, \citenamefont
  {Nishi}, \citenamefont {Oliveira},\ and\ \citenamefont
  {Vieira}}]{Barreto:2017xix}%
  \BibitemOpen
  \bibfield  {author} {\bibinfo {author} {\bibfnamefont {E.~R.}\ \bibnamefont
  {Barreto}}, \bibinfo {author} {\bibfnamefont {A.~G.}\ \bibnamefont {Dias}},
  \bibinfo {author} {\bibfnamefont {J.}~\bibnamefont {Leite}}, \bibinfo
  {author} {\bibfnamefont {C.~C.}\ \bibnamefont {Nishi}}, \bibinfo {author}
  {\bibfnamefont {R.~L.~N.}\ \bibnamefont {Oliveira}}, \ and\ \bibinfo {author}
  {\bibfnamefont {W.~C.}\ \bibnamefont {Vieira}},\ }\href {\doibase
  10.1103/PhysRevD.97.055047} {\bibfield  {journal} {\bibinfo  {journal} {Phys.
  Rev.}\ }\textbf {\bibinfo {volume} {D97}},\ \bibinfo {pages} {055047}
  (\bibinfo {year} {2018})},\ \Eprint {http://arxiv.org/abs/1709.09946}
  {arXiv:1709.09946 [hep-ph]} \BibitemShut {NoStop}%
\bibitem [{\citenamefont {Wei}\ and\ \citenamefont
  {Chong-Xing}(2017)}]{Wei:2017ago}%
  \BibitemOpen
  \bibfield  {author} {\bibinfo {author} {\bibfnamefont {M.}~\bibnamefont
  {Wei}}\ and\ \bibinfo {author} {\bibfnamefont {Y.}~\bibnamefont
  {Chong-Xing}},\ }\href {\doibase 10.1103/PhysRevD.95.035040} {\bibfield
  {journal} {\bibinfo  {journal} {Phys. Rev.}\ }\textbf {\bibinfo {volume}
  {D95}},\ \bibinfo {pages} {035040} (\bibinfo {year} {2017})},\ \Eprint
  {http://arxiv.org/abs/1702.01255} {arXiv:1702.01255 [hep-ph]} \BibitemShut
  {NoStop}%
\bibitem [{\citenamefont {Hue}\ \emph {et~al.}(2018)\citenamefont {Hue},
  \citenamefont {Ninh}, \citenamefont {Thuc},\ and\ \citenamefont
  {Dat}}]{Hue:2017lak}%
  \BibitemOpen
  \bibfield  {author} {\bibinfo {author} {\bibfnamefont {L.~T.}\ \bibnamefont
  {Hue}}, \bibinfo {author} {\bibfnamefont {L.~D.}\ \bibnamefont {Ninh}},
  \bibinfo {author} {\bibfnamefont {T.~T.}\ \bibnamefont {Thuc}}, \ and\
  \bibinfo {author} {\bibfnamefont {N.~T.~T.}\ \bibnamefont {Dat}},\ }\href
  {\doibase 10.1140/epjc/s10052-018-5589-3} {\bibfield  {journal} {\bibinfo
  {journal} {Eur. Phys. J.}\ }\textbf {\bibinfo {volume} {C78}},\ \bibinfo
  {pages} {128} (\bibinfo {year} {2018})},\ \Eprint
  {http://arxiv.org/abs/1708.09723} {arXiv:1708.09723 [hep-ph]} \BibitemShut
  {NoStop}%
\bibitem [{\citenamefont {Cogollo}\ \emph {et~al.}(2010)\citenamefont
  {Cogollo}, \citenamefont {Diniz},\ and\ \citenamefont
  {de~S.~Pires}}]{Cogollo:2010jw}%
  \BibitemOpen
  \bibfield  {author} {\bibinfo {author} {\bibfnamefont {D.}~\bibnamefont
  {Cogollo}}, \bibinfo {author} {\bibfnamefont {H.}~\bibnamefont {Diniz}}, \
  and\ \bibinfo {author} {\bibfnamefont {C.~A.}\ \bibnamefont {de~S.~Pires}},\
  }\href {\doibase 10.1016/j.physletb.2010.03.066} {\bibfield  {journal}
  {\bibinfo  {journal} {Phys. Lett.}\ }\textbf {\bibinfo {volume} {B687}},\
  \bibinfo {pages} {400} (\bibinfo {year} {2010})},\ \Eprint
  {http://arxiv.org/abs/1002.1944} {arXiv:1002.1944 [hep-ph]} \BibitemShut
  {NoStop}%
\bibitem [{\citenamefont {Cogollo}\ \emph {et~al.}(2008)\citenamefont
  {Cogollo}, \citenamefont {Diniz}, \citenamefont {de~S.~Pires},\ and\
  \citenamefont {Rodrigues~da Silva}}]{Cogollo:2008zc}%
  \BibitemOpen
  \bibfield  {author} {\bibinfo {author} {\bibfnamefont {D.}~\bibnamefont
  {Cogollo}}, \bibinfo {author} {\bibfnamefont {H.}~\bibnamefont {Diniz}},
  \bibinfo {author} {\bibfnamefont {C.~A.}\ \bibnamefont {de~S.~Pires}}, \ and\
  \bibinfo {author} {\bibfnamefont {P.~S.}\ \bibnamefont {Rodrigues~da
  Silva}},\ }\href {\doibase 10.1140/epjc/s10052-008-0749-5} {\bibfield
  {journal} {\bibinfo  {journal} {Eur. Phys. J.}\ }\textbf {\bibinfo {volume}
  {C58}},\ \bibinfo {pages} {455} (\bibinfo {year} {2008})},\ \Eprint
  {http://arxiv.org/abs/0806.3087} {arXiv:0806.3087 [hep-ph]} \BibitemShut
  {NoStop}%
\bibitem [{\citenamefont {Okada}\ \emph {et~al.}(2016)\citenamefont {Okada},
  \citenamefont {Okada},\ and\ \citenamefont {Orikasa}}]{Okada:2015bxa}%
  \BibitemOpen
  \bibfield  {author} {\bibinfo {author} {\bibfnamefont {H.}~\bibnamefont
  {Okada}}, \bibinfo {author} {\bibfnamefont {N.}~\bibnamefont {Okada}}, \ and\
  \bibinfo {author} {\bibfnamefont {Y.}~\bibnamefont {Orikasa}},\ }\href
  {\doibase 10.1103/PhysRevD.93.073006} {\bibfield  {journal} {\bibinfo
  {journal} {Phys. Rev.}\ }\textbf {\bibinfo {volume} {D93}},\ \bibinfo {pages}
  {073006} (\bibinfo {year} {2016})},\ \Eprint
  {http://arxiv.org/abs/1504.01204} {arXiv:1504.01204 [hep-ph]} \BibitemShut
  {NoStop}%
\bibitem [{\citenamefont {Vien}\ \emph {et~al.}(2019)\citenamefont {Vien},
  \citenamefont {Long},\ and\ \citenamefont
  {Carcamo~Hernandez}}]{Vien:2018otl}%
  \BibitemOpen
  \bibfield  {author} {\bibinfo {author} {\bibfnamefont {V.~V.}\ \bibnamefont
  {Vien}}, \bibinfo {author} {\bibfnamefont {H.~N.}\ \bibnamefont {Long}}, \
  and\ \bibinfo {author} {\bibfnamefont {A.~E.}\ \bibnamefont
  {Carcamo~Hernandez}},\ }\href {\doibase 10.1142/S0217732319500056} {\bibfield
   {journal} {\bibinfo  {journal} {Mod. Phys. Lett.}\ }\textbf {\bibinfo
  {volume} {A34}},\ \bibinfo {pages} {1950005} (\bibinfo {year} {2019})},\
  \Eprint {http://arxiv.org/abs/1812.07263} {arXiv:1812.07263 [hep-ph]}
  \BibitemShut {NoStop}%
\bibitem [{\citenamefont {Cárcamo~Hernández}\ \emph
  {et~al.}(2018)\citenamefont {Cárcamo~Hernández}, \citenamefont {Long},\
  and\ \citenamefont {Vien}}]{carcamoHernandez:2018iel}%
  \BibitemOpen
  \bibfield  {author} {\bibinfo {author} {\bibfnamefont {A.~E.}\ \bibnamefont
  {Cárcamo~Hernández}}, \bibinfo {author} {\bibfnamefont {H.~N.}\
  \bibnamefont {Long}}, \ and\ \bibinfo {author} {\bibfnamefont {V.~V.}\
  \bibnamefont {Vien}},\ }\href {\doibase 10.1140/epjc/s10052-018-6284-0}
  {\bibfield  {journal} {\bibinfo  {journal} {Eur. Phys. J.}\ }\textbf
  {\bibinfo {volume} {C78}},\ \bibinfo {pages} {804} (\bibinfo {year}
  {2018})},\ \Eprint {http://arxiv.org/abs/1803.01636} {arXiv:1803.01636
  [hep-ph]} \BibitemShut {NoStop}%
\bibitem [{\citenamefont {Nguyen}\ \emph {et~al.}(2018)\citenamefont {Nguyen},
  \citenamefont {Le}, \citenamefont {Hong},\ and\ \citenamefont
  {Hue}}]{Nguyen:2018rlb}%
  \BibitemOpen
  \bibfield  {author} {\bibinfo {author} {\bibfnamefont {T.~P.}\ \bibnamefont
  {Nguyen}}, \bibinfo {author} {\bibfnamefont {T.~T.}\ \bibnamefont {Le}},
  \bibinfo {author} {\bibfnamefont {T.~T.}\ \bibnamefont {Hong}}, \ and\
  \bibinfo {author} {\bibfnamefont {L.~T.}\ \bibnamefont {Hue}},\ }\href
  {\doibase 10.1103/PhysRevD.97.073003} {\bibfield  {journal} {\bibinfo
  {journal} {Phys. Rev.}\ }\textbf {\bibinfo {volume} {D97}},\ \bibinfo {pages}
  {073003} (\bibinfo {year} {2018})},\ \Eprint
  {http://arxiv.org/abs/1802.00429} {arXiv:1802.00429 [hep-ph]} \BibitemShut
  {NoStop}%
\bibitem [{\citenamefont {de~Sousa~Pires}\ \emph {et~al.}(2019)\citenamefont
  {de~Sousa~Pires}, \citenamefont {Ferreira De~Freitas}, \citenamefont {Shu},
  \citenamefont {Huang},\ and\ \citenamefont {Wagner
  Vasconcelos~Olegário}}]{Pires:2018kaj}%
  \BibitemOpen
  \bibfield  {author} {\bibinfo {author} {\bibfnamefont {C.~A.}\ \bibnamefont
  {de~Sousa~Pires}}, \bibinfo {author} {\bibfnamefont {F.}~\bibnamefont
  {Ferreira De~Freitas}}, \bibinfo {author} {\bibfnamefont {J.}~\bibnamefont
  {Shu}}, \bibinfo {author} {\bibfnamefont {L.}~\bibnamefont {Huang}}, \ and\
  \bibinfo {author} {\bibfnamefont {P.}~\bibnamefont {Wagner
  Vasconcelos~Olegário}},\ }\href {\doibase 10.1016/j.physletb.2019.134827}
  {\bibfield  {journal} {\bibinfo  {journal} {Phys. Lett.}\ }\textbf {\bibinfo
  {volume} {B797}},\ \bibinfo {pages} {134827} (\bibinfo {year} {2019})},\
  \Eprint {http://arxiv.org/abs/1812.10570} {arXiv:1812.10570 [hep-ph]}
  \BibitemShut {NoStop}%
\bibitem [{\citenamefont {Carcamo~Hernandez}\ \emph
  {et~al.}(2019{\natexlab{a}})\citenamefont {Carcamo~Hernandez}, \citenamefont
  {Perez-Julve},\ and\ \citenamefont
  {Hidalgo~Velasquez}}]{CarcamoHernandez:2019iwh}%
  \BibitemOpen
  \bibfield  {author} {\bibinfo {author} {\bibfnamefont {A.~E.}\ \bibnamefont
  {Carcamo~Hernandez}}, \bibinfo {author} {\bibfnamefont {N.~A.}\ \bibnamefont
  {Perez-Julve}}, \ and\ \bibinfo {author} {\bibfnamefont {Y.}~\bibnamefont
  {Hidalgo~Velasquez}},\ }\href {\doibase 10.1103/PhysRevD.100.095025}
  {\bibfield  {journal} {\bibinfo  {journal} {Phys. Rev.}\ }\textbf {\bibinfo
  {volume} {D100}},\ \bibinfo {pages} {095025} (\bibinfo {year}
  {2019}{\natexlab{a}})},\ \Eprint {http://arxiv.org/abs/1907.13083}
  {arXiv:1907.13083 [hep-ph]} \BibitemShut {NoStop}%
\bibitem [{\citenamefont {Carcamo~Hernandez}\ \emph
  {et~al.}(2019{\natexlab{b}})\citenamefont {Carcamo~Hernandez}, \citenamefont
  {Hidalgo~Velasquez},\ and\ \citenamefont
  {Perez-Julve}}]{CarcamoHernandez:2019vih}%
  \BibitemOpen
  \bibfield  {author} {\bibinfo {author} {\bibfnamefont {A.~E.}\ \bibnamefont
  {Carcamo~Hernandez}}, \bibinfo {author} {\bibfnamefont {Y.}~\bibnamefont
  {Hidalgo~Velasquez}}, \ and\ \bibinfo {author} {\bibfnamefont {N.~A.}\
  \bibnamefont {Perez-Julve}},\ }\href {\doibase
  10.1140/epjc/s10052-019-7325-z} {\bibfield  {journal} {\bibinfo  {journal}
  {Eur. Phys. J.}\ }\textbf {\bibinfo {volume} {C79}},\ \bibinfo {pages} {828}
  (\bibinfo {year} {2019}{\natexlab{b}})},\ \Eprint
  {http://arxiv.org/abs/1905.02323} {arXiv:1905.02323 [hep-ph]} \BibitemShut
  {NoStop}%
\bibitem [{\citenamefont {Cárcamo~Hernández}\ \emph
  {et~al.}(2020)\citenamefont {Cárcamo~Hernández}, \citenamefont {Hue},
  \citenamefont {Kovalenko},\ and\ \citenamefont
  {Long}}]{CarcamoHernandez:2020pnh}%
  \BibitemOpen
  \bibfield  {author} {\bibinfo {author} {\bibfnamefont {A.~E.}\ \bibnamefont
  {Cárcamo~Hernández}}, \bibinfo {author} {\bibfnamefont {L.~T.}\
  \bibnamefont {Hue}}, \bibinfo {author} {\bibfnamefont {S.}~\bibnamefont
  {Kovalenko}}, \ and\ \bibinfo {author} {\bibfnamefont {H.~N.}\ \bibnamefont
  {Long}},\ }\href@noop {} {\  (\bibinfo {year} {2020})},\ \Eprint
  {http://arxiv.org/abs/2001.01748} {arXiv:2001.01748 [hep-ph]} \BibitemShut
  {NoStop}%
\bibitem [{\citenamefont {Meirose}\ and\ \citenamefont
  {Nepomuceno}(2011)}]{Meirose:2011cs}%
  \BibitemOpen
  \bibfield  {author} {\bibinfo {author} {\bibfnamefont {B.}~\bibnamefont
  {Meirose}}\ and\ \bibinfo {author} {\bibfnamefont {A.~A.}\ \bibnamefont
  {Nepomuceno}},\ }\href {\doibase 10.1103/PhysRevD.84.055002} {\bibfield
  {journal} {\bibinfo  {journal} {Phys. Rev.}\ }\textbf {\bibinfo {volume}
  {D84}},\ \bibinfo {pages} {055002} (\bibinfo {year} {2011})},\ \Eprint
  {http://arxiv.org/abs/1105.6299} {arXiv:1105.6299 [hep-ph]} \BibitemShut
  {NoStop}%
\bibitem [{\citenamefont {Coutinho}\ \emph {et~al.}(2013)\citenamefont
  {Coutinho}, \citenamefont {Salustino~Guimarães},\ and\ \citenamefont
  {Nepomuceno}}]{Coutinho:2013lta}%
  \BibitemOpen
  \bibfield  {author} {\bibinfo {author} {\bibfnamefont {Y.~A.}\ \bibnamefont
  {Coutinho}}, \bibinfo {author} {\bibfnamefont {V.}~\bibnamefont
  {Salustino~Guimarães}}, \ and\ \bibinfo {author} {\bibfnamefont {A.~A.}\
  \bibnamefont {Nepomuceno}},\ }\href {\doibase 10.1103/PhysRevD.87.115014}
  {\bibfield  {journal} {\bibinfo  {journal} {Phys. Rev.}\ }\textbf {\bibinfo
  {volume} {D87}},\ \bibinfo {pages} {115014} (\bibinfo {year} {2013})},\
  \Eprint {http://arxiv.org/abs/1304.7907} {arXiv:1304.7907 [hep-ph]}
  \BibitemShut {NoStop}%
\bibitem [{\citenamefont {Nepomuceno}\ \emph {et~al.}(2016)\citenamefont
  {Nepomuceno}, \citenamefont {Meirose},\ and\ \citenamefont
  {Eccard}}]{Nepomuceno:2016jyr}%
  \BibitemOpen
  \bibfield  {author} {\bibinfo {author} {\bibfnamefont {A.}~\bibnamefont
  {Nepomuceno}}, \bibinfo {author} {\bibfnamefont {B.}~\bibnamefont {Meirose}},
  \ and\ \bibinfo {author} {\bibfnamefont {F.}~\bibnamefont {Eccard}},\ }\href
  {\doibase 10.1103/PhysRevD.94.055020} {\bibfield  {journal} {\bibinfo
  {journal} {Phys. Rev.}\ }\textbf {\bibinfo {volume} {D94}},\ \bibinfo {pages}
  {055020} (\bibinfo {year} {2016})},\ \Eprint
  {http://arxiv.org/abs/1604.07471} {arXiv:1604.07471 [hep-ph]} \BibitemShut
  {NoStop}%
\bibitem [{\citenamefont {Nepomuceno}\ and\ \citenamefont
  {Meirose}(2020)}]{Nepomuceno:2019eaz}%
  \BibitemOpen
  \bibfield  {author} {\bibinfo {author} {\bibfnamefont {A.}~\bibnamefont
  {Nepomuceno}}\ and\ \bibinfo {author} {\bibfnamefont {B.}~\bibnamefont
  {Meirose}},\ }\href {\doibase 10.1103/PhysRevD.101.035017} {\bibfield
  {journal} {\bibinfo  {journal} {Phys. Rev.}\ }\textbf {\bibinfo {volume}
  {D101}},\ \bibinfo {pages} {035017} (\bibinfo {year} {2020})},\ \Eprint
  {http://arxiv.org/abs/1911.12783} {arXiv:1911.12783 [hep-ph]} \BibitemShut
  {NoStop}%
\bibitem [{\citenamefont {Borges}\ and\ \citenamefont
  {Ramos}(2016)}]{Borges:2016nne}%
  \BibitemOpen
  \bibfield  {author} {\bibinfo {author} {\bibfnamefont {J.~S.}\ \bibnamefont
  {Borges}}\ and\ \bibinfo {author} {\bibfnamefont {R.~O.}\ \bibnamefont
  {Ramos}},\ }\href {\doibase 10.1140/epjc/s10052-016-4168-8} {\bibfield
  {journal} {\bibinfo  {journal} {Eur. Phys. J.}\ }\textbf {\bibinfo {volume}
  {C76}},\ \bibinfo {pages} {344} (\bibinfo {year} {2016})},\ \Eprint
  {http://arxiv.org/abs/1602.08165} {arXiv:1602.08165 [hep-ph]} \BibitemShut
  {NoStop}%
\bibitem [{\citenamefont {Ponce}\ \emph
  {et~al.}(2002{\natexlab{a}})\citenamefont {Ponce}, \citenamefont {Florez},\
  and\ \citenamefont {Sanchez}}]{Ponce:2001jn}%
  \BibitemOpen
  \bibfield  {author} {\bibinfo {author} {\bibfnamefont {W.~A.}\ \bibnamefont
  {Ponce}}, \bibinfo {author} {\bibfnamefont {J.~B.}\ \bibnamefont {Florez}}, \
  and\ \bibinfo {author} {\bibfnamefont {L.~A.}\ \bibnamefont {Sanchez}},\
  }\href {\doibase 10.1142/S0217751X02005815} {\bibfield  {journal} {\bibinfo
  {journal} {Int. J. Mod. Phys.}\ }\textbf {\bibinfo {volume} {A17}},\ \bibinfo
  {pages} {643} (\bibinfo {year} {2002}{\natexlab{a}})},\ \Eprint
  {http://arxiv.org/abs/hep-ph/0103100} {arXiv:hep-ph/0103100 [hep-ph]}
  \BibitemShut {NoStop}%
\bibitem [{\citenamefont {Ponce}\ \emph
  {et~al.}(2002{\natexlab{b}})\citenamefont {Ponce}, \citenamefont {Giraldo},\
  and\ \citenamefont {Sanchez}}]{Ponce:2002fv}%
  \BibitemOpen
  \bibfield  {author} {\bibinfo {author} {\bibfnamefont {W.~A.}\ \bibnamefont
  {Ponce}}, \bibinfo {author} {\bibfnamefont {Y.}~\bibnamefont {Giraldo}}, \
  and\ \bibinfo {author} {\bibfnamefont {L.~A.}\ \bibnamefont {Sanchez}},\
  }\bibfield  {booktitle} {\emph {\bibinfo {booktitle} {{Proceedings, 8th
  Mexican Workshop on Particles and Fields: Zacatecas, Mexico, November 14-20,
  2001}}},\ }\href {\doibase 10.1063/1.1489776} {\bibfield  {journal} {\bibinfo
   {journal} {AIP Conf. Proc.}\ }\textbf {\bibinfo {volume} {623}},\ \bibinfo
  {pages} {341} (\bibinfo {year} {2002}{\natexlab{b}})},\ \Eprint
  {http://arxiv.org/abs/hep-ph/0201133} {arXiv:hep-ph/0201133 [hep-ph]}
  \BibitemShut {NoStop}%
\bibitem [{\citenamefont {Anderson}\ and\ \citenamefont
  {Sher}(2005)}]{Anderson:2005ab}%
  \BibitemOpen
  \bibfield  {author} {\bibinfo {author} {\bibfnamefont {D.~L.}\ \bibnamefont
  {Anderson}}\ and\ \bibinfo {author} {\bibfnamefont {M.}~\bibnamefont
  {Sher}},\ }\href {\doibase 10.1103/PhysRevD.72.095014} {\bibfield  {journal}
  {\bibinfo  {journal} {Phys. Rev.}\ }\textbf {\bibinfo {volume} {D72}},\
  \bibinfo {pages} {095014} (\bibinfo {year} {2005})},\ \Eprint
  {http://arxiv.org/abs/hep-ph/0509200} {arXiv:hep-ph/0509200 [hep-ph]}
  \BibitemShut {NoStop}%
\bibitem [{\citenamefont {Cabarcas}\ \emph {et~al.}(2014)\citenamefont
  {Cabarcas}, \citenamefont {Duarte},\ and\ \citenamefont
  {Rodriguez}}]{Cabarcas:2013jba}%
  \BibitemOpen
  \bibfield  {author} {\bibinfo {author} {\bibfnamefont {J.~M.}\ \bibnamefont
  {Cabarcas}}, \bibinfo {author} {\bibfnamefont {J.}~\bibnamefont {Duarte}}, \
  and\ \bibinfo {author} {\bibfnamefont {J.~A.}\ \bibnamefont {Rodriguez}},\
  }\href {\doibase 10.1142/S0217751X14500158} {\bibfield  {journal} {\bibinfo
  {journal} {Int. J. Mod. Phys.}\ }\textbf {\bibinfo {volume} {A29}},\ \bibinfo
  {pages} {1450015} (\bibinfo {year} {2014})},\ \Eprint
  {http://arxiv.org/abs/1310.1407} {arXiv:1310.1407 [hep-ph]} \BibitemShut
  {NoStop}%
\bibitem [{\citenamefont {Long}(1996{\natexlab{b}})}]{Hoang:1996gi}%
  \BibitemOpen
  \bibfield  {author} {\bibinfo {author} {\bibfnamefont {H.~N.}\ \bibnamefont
  {Long}},\ }\href {\doibase 10.1103/PhysRevD.54.4691} {\bibfield  {journal}
  {\bibinfo  {journal} {Phys. Rev.}\ }\textbf {\bibinfo {volume} {D54}},\
  \bibinfo {pages} {4691} (\bibinfo {year} {1996}{\natexlab{b}})},\ \Eprint
  {http://arxiv.org/abs/hep-ph/9607439} {arXiv:hep-ph/9607439 [hep-ph]}
  \BibitemShut {NoStop}%
\bibitem [{\citenamefont {Catano}\ \emph {et~al.}(2012)\citenamefont {Catano},
  \citenamefont {Martinez},\ and\ \citenamefont {Ochoa}}]{Catano:2012kw}%
  \BibitemOpen
  \bibfield  {author} {\bibinfo {author} {\bibfnamefont {M.~E.}\ \bibnamefont
  {Catano}}, \bibinfo {author} {\bibfnamefont {R.}~\bibnamefont {Martinez}}, \
  and\ \bibinfo {author} {\bibfnamefont {F.}~\bibnamefont {Ochoa}},\ }\href
  {\doibase 10.1103/PhysRevD.86.073015} {\bibfield  {journal} {\bibinfo
  {journal} {Phys. Rev.}\ }\textbf {\bibinfo {volume} {D86}},\ \bibinfo {pages}
  {073015} (\bibinfo {year} {2012})},\ \Eprint {http://arxiv.org/abs/1206.1966}
  {arXiv:1206.1966 [hep-ph]} \BibitemShut {NoStop}%
\bibitem [{\citenamefont {Dong}\ \emph {et~al.}(2006)\citenamefont {Dong},
  \citenamefont {Long}, \citenamefont {Nhung},\ and\ \citenamefont
  {Soa}}]{Dong:2006mg}%
  \BibitemOpen
  \bibfield  {author} {\bibinfo {author} {\bibfnamefont {P.~V.}\ \bibnamefont
  {Dong}}, \bibinfo {author} {\bibfnamefont {H.~N.}\ \bibnamefont {Long}},
  \bibinfo {author} {\bibfnamefont {D.~T.}\ \bibnamefont {Nhung}}, \ and\
  \bibinfo {author} {\bibfnamefont {D.~V.}\ \bibnamefont {Soa}},\ }\href
  {\doibase 10.1103/PhysRevD.73.035004} {\bibfield  {journal} {\bibinfo
  {journal} {Phys. Rev.}\ }\textbf {\bibinfo {volume} {D73}},\ \bibinfo {pages}
  {035004} (\bibinfo {year} {2006})},\ \Eprint
  {http://arxiv.org/abs/hep-ph/0601046} {arXiv:hep-ph/0601046 [hep-ph]}
  \BibitemShut {NoStop}%
\bibitem [{\citenamefont {Dong}\ and\ \citenamefont
  {Long}(2008)}]{Dong:2008ya}%
  \BibitemOpen
  \bibfield  {author} {\bibinfo {author} {\bibfnamefont {P.~V.}\ \bibnamefont
  {Dong}}\ and\ \bibinfo {author} {\bibfnamefont {H.~N.}\ \bibnamefont
  {Long}},\ }\href {\doibase 10.1155/2008/739492} {\bibfield  {journal}
  {\bibinfo  {journal} {Adv. High Energy Phys.}\ }\textbf {\bibinfo {volume}
  {2008}},\ \bibinfo {pages} {739492} (\bibinfo {year} {2008})},\ \Eprint
  {http://arxiv.org/abs/0804.3239} {arXiv:0804.3239 [hep-ph]} \BibitemShut
  {NoStop}%
\bibitem [{\citenamefont {Berenstein}\ \emph {et~al.}(2009)\citenamefont
  {Berenstein}, \citenamefont {Martinez}, \citenamefont {Ochoa},\ and\
  \citenamefont {Pinansky}}]{Berenstein:2008xg}%
  \BibitemOpen
  \bibfield  {author} {\bibinfo {author} {\bibfnamefont {D.}~\bibnamefont
  {Berenstein}}, \bibinfo {author} {\bibfnamefont {R.}~\bibnamefont
  {Martinez}}, \bibinfo {author} {\bibfnamefont {F.}~\bibnamefont {Ochoa}}, \
  and\ \bibinfo {author} {\bibfnamefont {S.}~\bibnamefont {Pinansky}},\ }\href
  {\doibase 10.1103/PhysRevD.79.095005} {\bibfield  {journal} {\bibinfo
  {journal} {Phys. Rev.}\ }\textbf {\bibinfo {volume} {D79}},\ \bibinfo {pages}
  {095005} (\bibinfo {year} {2009})},\ \Eprint {http://arxiv.org/abs/0807.1126}
  {arXiv:0807.1126 [hep-ph]} \BibitemShut {NoStop}%
\bibitem [{\citenamefont {Martinez}\ and\ \citenamefont
  {Ochoa}(2014)}]{Martinez:2014lta}%
  \BibitemOpen
  \bibfield  {author} {\bibinfo {author} {\bibfnamefont {R.}~\bibnamefont
  {Martinez}}\ and\ \bibinfo {author} {\bibfnamefont {F.}~\bibnamefont
  {Ochoa}},\ }\href {\doibase 10.1103/PhysRevD.90.015028} {\bibfield  {journal}
  {\bibinfo  {journal} {Phys. Rev.}\ }\textbf {\bibinfo {volume} {D90}},\
  \bibinfo {pages} {015028} (\bibinfo {year} {2014})},\ \Eprint
  {http://arxiv.org/abs/1405.4566} {arXiv:1405.4566 [hep-ph]} \BibitemShut
  {NoStop}%
\bibitem [{\citenamefont {Ky}\ \emph {et~al.}(2000)\citenamefont {Ky},
  \citenamefont {Long},\ and\ \citenamefont {Van~Soa}}]{Ky:2000ku}%
  \BibitemOpen
  \bibfield  {author} {\bibinfo {author} {\bibfnamefont {N.~A.}\ \bibnamefont
  {Ky}}, \bibinfo {author} {\bibfnamefont {H.~N.}\ \bibnamefont {Long}}, \ and\
  \bibinfo {author} {\bibfnamefont {D.}~\bibnamefont {Van~Soa}},\ }\href
  {\doibase 10.1016/S0370-2693(00)00696-1} {\bibfield  {journal} {\bibinfo
  {journal} {Phys. Lett.}\ }\textbf {\bibinfo {volume} {B486}},\ \bibinfo
  {pages} {140} (\bibinfo {year} {2000})},\ \Eprint
  {http://arxiv.org/abs/hep-ph/0007010} {arXiv:hep-ph/0007010 [hep-ph]}
  \BibitemShut {NoStop}%
\bibitem [{\citenamefont {Kelso}\ \emph {et~al.}(2014)\citenamefont {Kelso},
  \citenamefont {Pinheiro}, \citenamefont {Queiroz},\ and\ \citenamefont
  {Shepherd}}]{Kelso:2013zfa}%
  \BibitemOpen
  \bibfield  {author} {\bibinfo {author} {\bibfnamefont {C.}~\bibnamefont
  {Kelso}}, \bibinfo {author} {\bibfnamefont {P.~R.~D.}\ \bibnamefont
  {Pinheiro}}, \bibinfo {author} {\bibfnamefont {F.~S.}\ \bibnamefont
  {Queiroz}}, \ and\ \bibinfo {author} {\bibfnamefont {W.}~\bibnamefont
  {Shepherd}},\ }\href {\doibase 10.1140/epjc/s10052-014-2808-4} {\bibfield
  {journal} {\bibinfo  {journal} {Eur. Phys. J.}\ }\textbf {\bibinfo {volume}
  {C74}},\ \bibinfo {pages} {2808} (\bibinfo {year} {2014})},\ \Eprint
  {http://arxiv.org/abs/1312.0051} {arXiv:1312.0051 [hep-ph]} \BibitemShut
  {NoStop}%
\bibitem [{\citenamefont {Binh}\ \emph {et~al.}(2015)\citenamefont {Binh},
  \citenamefont {Huong},\ and\ \citenamefont {Long}}]{Binh:2015cba}%
  \BibitemOpen
  \bibfield  {author} {\bibinfo {author} {\bibfnamefont {D.~T.}\ \bibnamefont
  {Binh}}, \bibinfo {author} {\bibfnamefont {D.~T.}\ \bibnamefont {Huong}}, \
  and\ \bibinfo {author} {\bibfnamefont {H.~N.}\ \bibnamefont {Long}},\ }\href
  {\doibase 10.7868/S004445101512007X, 10.1134/S1063776115120109} {\bibfield
  {journal} {\bibinfo  {journal} {Zh. Eksp. Teor. Fiz.}\ }\textbf {\bibinfo
  {volume} {148}},\ \bibinfo {pages} {1115} (\bibinfo {year} {2015})},\
  \bibinfo {note} {[J. Exp. Theor. Phys.121,no.6,976(2015)]},\ \Eprint
  {http://arxiv.org/abs/1504.03510} {arXiv:1504.03510 [hep-ph]} \BibitemShut
  {NoStop}%
\bibitem [{\citenamefont {Cogollo}(2017)}]{Cogollo:2017foz}%
  \BibitemOpen
  \bibfield  {author} {\bibinfo {author} {\bibfnamefont {D.}~\bibnamefont
  {Cogollo}},\ }\href@noop {} {\  (\bibinfo {year} {2017})},\ \Eprint
  {http://arxiv.org/abs/1706.00397} {arXiv:1706.00397 [hep-ph]} \BibitemShut
  {NoStop}%
\bibitem [{\citenamefont {De~Conto}\ and\ \citenamefont
  {Pleitez}(2017)}]{DeConto:2016ith}%
  \BibitemOpen
  \bibfield  {author} {\bibinfo {author} {\bibfnamefont {G.}~\bibnamefont
  {De~Conto}}\ and\ \bibinfo {author} {\bibfnamefont {V.}~\bibnamefont
  {Pleitez}},\ }\href {\doibase 10.1007/JHEP05(2017)104} {\bibfield  {journal}
  {\bibinfo  {journal} {JHEP}\ }\textbf {\bibinfo {volume} {05}},\ \bibinfo
  {pages} {104} (\bibinfo {year} {2017})},\ \Eprint
  {http://arxiv.org/abs/1603.09691} {arXiv:1603.09691 [hep-ph]} \BibitemShut
  {NoStop}%
\bibitem [{\citenamefont {de~Jesus}\ \emph {et~al.}(2020)\citenamefont
  {de~Jesus}, \citenamefont {Kovalenko}, \citenamefont {Queiroz}, \citenamefont
  {Pires},\ and\ \citenamefont {Villamizar}}]{deJesus:2020ngn}%
  \BibitemOpen
  \bibfield  {author} {\bibinfo {author} {\bibfnamefont {A.~S.}\ \bibnamefont
  {de~Jesus}}, \bibinfo {author} {\bibfnamefont {S.}~\bibnamefont {Kovalenko}},
  \bibinfo {author} {\bibfnamefont {F.~S.}\ \bibnamefont {Queiroz}}, \bibinfo
  {author} {\bibfnamefont {C.~A. d.~S.}\ \bibnamefont {Pires}}, \ and\ \bibinfo
  {author} {\bibfnamefont {Y.~S.}\ \bibnamefont {Villamizar}},\ }\href@noop {}
  {\  (\bibinfo {year} {2020})},\ \Eprint {http://arxiv.org/abs/2003.06440}
  {arXiv:2003.06440 [hep-ph]} \BibitemShut {NoStop}%
\bibitem [{\citenamefont {Cid~Vidal}\ \emph {et~al.}(2018)\citenamefont
  {Cid~Vidal} \emph {et~al.}}]{CidVidal:2018eel}%
  \BibitemOpen
  \bibfield  {author} {\bibinfo {author} {\bibfnamefont {X.}~\bibnamefont
  {Cid~Vidal}} \emph {et~al.} (\bibinfo {collaboration} {Working Group 3}),\
  }\href@noop {} {\  (\bibinfo {year} {2018})},\ \Eprint
  {http://arxiv.org/abs/1812.07831} {arXiv:1812.07831 [hep-ph]} \BibitemShut
  {NoStop}%
\bibitem [{\citenamefont {Thamm}\ \emph {et~al.}(2015)\citenamefont {Thamm},
  \citenamefont {Torre},\ and\ \citenamefont {Wulzer}}]{Thamm:2015zwa}%
  \BibitemOpen
  \bibfield  {author} {\bibinfo {author} {\bibfnamefont {A.}~\bibnamefont
  {Thamm}}, \bibinfo {author} {\bibfnamefont {R.}~\bibnamefont {Torre}}, \ and\
  \bibinfo {author} {\bibfnamefont {A.}~\bibnamefont {Wulzer}},\ }\href
  {\doibase 10.1007/JHEP07(2015)100} {\bibfield  {journal} {\bibinfo  {journal}
  {JHEP}\ }\textbf {\bibinfo {volume} {07}},\ \bibinfo {pages} {100} (\bibinfo
  {year} {2015})},\ \Eprint {http://arxiv.org/abs/1502.01701} {arXiv:1502.01701
  [hep-ph]} \BibitemShut {NoStop}%
\bibitem [{\citenamefont {Carcamo~Hernandez}\ \emph {et~al.}(2020)\citenamefont
  {Carcamo~Hernandez}, \citenamefont {Velasquez}, \citenamefont {Kovalenko},
  \citenamefont {Long}, \citenamefont {Perez-Julve},\ and\ \citenamefont
  {Vien}}]{CarcamoHernandez:2020pxw}%
  \BibitemOpen
  \bibfield  {author} {\bibinfo {author} {\bibfnamefont {A.~E.}\ \bibnamefont
  {Carcamo~Hernandez}}, \bibinfo {author} {\bibfnamefont {Y.~H.}\ \bibnamefont
  {Velasquez}}, \bibinfo {author} {\bibfnamefont {S.}~\bibnamefont
  {Kovalenko}}, \bibinfo {author} {\bibfnamefont {H.~N.}\ \bibnamefont {Long}},
  \bibinfo {author} {\bibfnamefont {N.~A.}\ \bibnamefont {Perez-Julve}}, \ and\
  \bibinfo {author} {\bibfnamefont {V.~V.}\ \bibnamefont {Vien}},\ }\href@noop
  {} {\  (\bibinfo {year} {2020})},\ \Eprint {http://arxiv.org/abs/2002.07347}
  {arXiv:2002.07347 [hep-ph]} \BibitemShut {NoStop}%
\end{thebibliography}%

\end{document}